\documentclass[]{aastex701}

\usepackage{booktabs}
\usepackage{graphicx}
\usepackage{longtable}
\usepackage{threeparttablex}
\usepackage{xcolor}{}
\begin{document}

\title{A Study of HH 270 with the James Webb Space Telescope}

\author[orcid=0009-0007-1261-4893]{A. N. Ortiz Capeles}
\affiliation{Universidad de Puerto Rico - Río Piedras \\
Natural Sciences Faculty, Physics Department \\
San Juan, PR 00925-2537, USA}
\email[show]{alexander.ortiz12@upr.edu}

\author[orcid=0000-0002-6296-8960]{A. Noriega-Crespo}
\affiliation{Space Telescope Science Institute \\
3700 San Martin Drive \\
Baltimore, MD 21218, USA}
\email[show]{anoriega@stsci.edu}

\author[orcid=0000-0002-0835-1126]{A. C. Raga}
\affiliation{Universidad Nacional Autónoma de México \\
Circuito Exterior S/N, Ciudad Universitaria \\
Apartado Postal 70-543, Del. Coyoacán, C.P. 04510, México D.F.}
\email{raga@nucleares.unam.mx}

\author[orcid=0000-0001-7943-9961]{M. E. Lebrón}
\affiliation{Universidad de Puerto Rico - Río Piedras \\
General Studies Faculty, Physical Sciences Department \\
San Juan, PR 00925-2532, USA}
\email[show]{mayra.lebron3@upr.edu}

\author[orcid=0000-0001-5653-7817]{H. Arce}
\affiliation{Yale University - Astronomy Department \\
260 Whitney Avenue \\
New Haven, CT 06511 USA}
\email[show]{hector.arce@yale.edu}

\author[orcid=0000-0003-0759-4991]{J. L. Morales Ortiz}
\affiliation{Universidad de Puerto Rico - Río Piedras \\
General Studies Faculty, Physical Sciences Department \\
San Juan, PR 00925-2532, USA}
\email[show]{jorge.morales15@upr.edu}

\author[orcid=0000-0003-0759-4991]{C. A. Pantoja}
\affiliation{Universidad de Puerto Rico - Río Piedras \\
Natural Sciences Faculty, Physics Department \\
San Juan, PR 00925-2537, USA}
\email[show]{carmen.pantoja1@upr.edu}

\begin{abstract}
We present a study of the Herbig-Haro object HH 270 based on observations from the James Webb Space Telescope (JWST), Subaru Telescope, and Atacama Large Millimeter/submillimeter Array (ALMA). 
High-resolution infrared images of H$_2$ and CO were obtained with the NIRCam instrument (JWST) using the F212N (2.12 $\mu$m) and F460M (4.60 $\mu$m) filters, revealing a previously unseen collimated protostellar jet closer to the source, in addition to the very well defined bipolar cavities carved by the outflow.
Newly identified knots associated with the jet were also detected. 
Ground-based optical images in the H$\alpha$ (660 nm) emission line, alongside millimeter spectral observations of the (2-1) transition of $^{12}$CO, $^{13}$CO, and C$^{18}$O, further enrich the analysis. 
The Subaru images show a connection between the optical outflow in H$\alpha$ and the protostellar jet observed in the infrared. 
ALMA CO observations trace the kinematics of the entrained molecular gas in the protostellar outflow and reveal the dense, slow-moving material distributed around the driving source, HH270VLA1. 
These multi-wavelength observations show evidence of the interaction between the shock-excited jet emission and the molecular outflow seen at optical, infrared and radio wavelengths, which provides a detailed view of the complex structure and dynamics of HH 270.
\end{abstract}


\section{Introduction}\label{sec:intro}
Herbig-Haro (HH) objects are shock-excited nebulae formed by the interaction between collimated jets from young stellar objects (YSOs) and the surrounding interstellar medium (ISM). 
These jets, ejected during the early stages of star formation, collide with ambient gas and dust, producing shock fronts that emit characteristic spectral lines across a broad range of wavelengths. 
The morphology of HH objects can vary widely, from chaotic, fragmented structures to well-defined jets or bow shocks. 
HH objects are critical probes for studying the interaction between stellar jets and the ISM, revealing the dynamic and complex nature of star formation, where these interactions drive the morphology and evolution of the jet-ISM system(\citealt{2001ARA&A..39..403R}).

A particular active region where such phenomena are found is the L1617 molecular cloud, located within the Orion B complex at a distance of $\sim$430 pc from the Sun (\citealt{2022ApJ...925...39T}). 
This region contains multiple dense cores and hosts several HH objects, including HH 110, 111, 112, 121, and the subject of this study, HH 270 (\citealt{2005A&A...437..169W}; \citealt{2008hsf1.book..782R}). 
The L1617 cloud is notable for its dynamic environment, shaped by interactions between protostellar outflows within the molecular cloud. 
These interactions give rise to diverse jet morphologies, which in turn influence subsequent star formation.

In this context, the HH 270/110 system presents a compelling case of jet deflection, likely caused by a grazing collision between the HH 270 jet and a dense molecular clump, resulting in the formation of the adjacent HH 110 flow (\citealt{1996A&A...311..989R}). 
This system has been the subject of numerous multi-wavelength studies, each contributing to our understanding of jet-cloud dynamics in star-forming environments (see \S \ref{sec:background}).

This paper aims to build upon those studies by providing new high-resolution observations of HH 270 with the NIRCam instrument (\citealt{2023PASP..135b8001R}) and the Atacama Large Millimeter/submillimeter Array (ALMA). 
\S \ref{sec:background} provides a detailed background on HH 270 and summarizes relevant prior observations. 
\S \ref{sec:Observations} outlines the NIRCam, ALMA and Subaru datasets used. 
In \S \ref{sec:Results}, we present our observational results, including the procedure for identifying jet knots, and the derived parameters from infrared (IR) and radio observations. 
\S \ref{sec:Discussion} discusses the cavity structure and kinematic features identified in the NIRCam and ALMA data, and compares them with previous optical studies. 
We summarize our conclusions and highlight future directions in \S \ref{sec:Conclusion}.

\section{Source Background} \label{sec:background}
Observations across multiple wavelengths reveal how the HH 270 jet interacts with its environment in the L1617 molecular cloud. 
Early optical imaging and spectroscopy (\citealt{1996A&A...311..989R}) demonstrated that the HH 270/110 deflection produced a noticeable brightening and altered the jet's trajectory, suggesting that jet-cloud collisions can play a major role in shaping protostellar outflows. 

Near-IR observations (\citealt{1996ApJ...462..804N}) confirmed the interaction scenario revealing bright H$_2$ emission along the deflected jet path with a collision angle of $\sim$60 degrees. 
While various theoretical models were proposed to explain the morphology and energetics of the system, they struggled to account for the observed sudden brightening at the deflection point, leaving this feature as an unresolved aspect of the jet-cloud interaction. 

Subsequent radio continuum observations with the VLA (\citealt{1998RMxAA..34...69R}) provided a confirmation that HH 110 is not an independent outflow, but a deflected portion of the HH 270 jet. 
These observations identified a compact radio source, VLA1 (now referred to as HH270VLA1), as the driving protostar of the entire outflow system. 
In addition, NH$_{3}$ observations (\citealt{2011A&A...527A..41S}), revealed that the molecular gas responsible for the jet's deflection is both dense and turbulent, supporting the interpretation that the environment shapes the jet's path.

Millimeter observations by \cite{2006ApJ...648..504C} showed that HH270VLA1 is a $\sim$1.6 M$_{\odot}$ protostar embedded in a flattened core perpendicular to the jet, with evidence for infall motions in the observed $^{13}$CO spectrum. 
Their study confirmed that HH270VLA1 is the driving source of the HH 270 outflow and established that HH 270 IRS and VLA1 are separate objects, potentially part of a protobinary system. 

This interpretation gained further support from the VANDAM (VLA/ALMA Nascent Disk and Multiplicity) survey (\citealt{2020ApJ...890..130T}; \citeyear{2022ApJ...925...39T}). 
This study confirmed that HH270VLA1 is a Class 0 protostar with two closely placed components (VLA1-A and VLA1-B) separated by $\sim$100 au and a third non-thermal source. 
This last source is likely associated with shock activity rather than being a true protostellar companion. 
These findings highlighted the complexity of distinguishing between multiplicity and outflow-related structures in deeply embedded systems and reinforced the challenges of interpreting kinematics and morphology in such environments.

New knots associated with the HH 270 jet were identified by \cite{2012AJ....143..106K}, using observations from the Subaru Telescope, the New Technology Telescope (NTT), and the Hubble Space Telescope (HST). 
Their proper motions were analyzed to provide details that revealed the HH 270 jet had undergone past axis shifts that may have been a crucial factor in the jet’s deflection, resulting in the formation of HH 110 (\citealt{2012AJ....143..106K}). 
The discovery of new HH objects near HH 270, also by \cite{2012AJ....143..106K}, reinforced the view that past interactions between the jet and the surrounding environment played a critical role in shaping both the HH 270 and HH 110 outflows.

Most recently, JWST NIRCam observations (\citealt{2025ApJ...981...95R}) revealed the detailed morphology of HH 270's bipolar cavity, detecting for the first time a highly collimated redshifted jet extending northeast from HH270VLA1. 
These observations, in the 4.60 $\mu$m (F460M) and 2.12 $\mu$m (F212N) bands, traced pure-rotational and vibrational H$_2$ lines, offering unprecedented spatial resolution. 
The authors proposed that the morphology and deflection angle observed in HH 110 could be explained by a jet-shear interaction, where the jet propagates through a shearing ambient medium instead of  simply colliding with a dense cloud, as previously thought. 
Their numerical simulations reproduced the observed internal working surfaces and proper motions, presenting an alternative view of jet-environment dynamics in the system.

All together, these studies emphasize HH 270 as a key case of study for understanding how protostellar jets evolve in turbulent, structured environments. 
They reveal that jet deflection, turbulence, and shear flows must all be considered to fully interpret the observed features of the HH 270/110 system.

\section{Observations and Analysis} \label{sec:Observations}
\subsection{JWST NIRCam Observations}
As outlined in \cite{2025ApJ...981...95R}, the NIRCam data of the HH 270/110 system were obtained under the Guaranteed Time Observations (GTO) 1293 program (\citealt{2017jwst.prop.1293N}). 
Observations were carried out using the F460M and F212N filters, employing a SHALLOW4 readout pattern with 4 groups and 3 dithers across a 3 $\times$ 2 mosaic with 15\% overlap. 
Each tile had an exposure time of 612 seconds, totaling 3672 seconds for the entire mosaic. 
The final mosaicked image covers $\sim0.7 \arcmin \times 0.8 \arcmin$ with an angular resolution of $\sim$126 mas. 
The images used are Level 3 pipeline products from STScI and no further processing has been applied. 
For the mosaic generation, the calibration software version used was \texttt{1.12.5}, with the Calibration Reference Data System (CRDS) software version \texttt{11.17.6}. 
At the time of this publication, newer versions may be available through the STScI MAST Portal.

The F460M filter traces H$_2$ emission from the 0-0 S(9) 4.69 $\mu$m pure-rotational line and possibly includes the $^{12}$CO fundamental R-branch ($\sim$4.6-4.8 $\mu$m), both excellent indicators of shocked gas in young stellar outflows (e.g., \citealt{2023Natur.622...48R}). 
The F212N narrowband filter is centered on the 1-0 S(1) 2.12 $\mu$m H$_2$ line, a reliable tracer of shock excitation in protostellar environments. The NIRCam data used in this paper can be found in MAST: \dataset[10.17909/51hc-z182]{http://dx.doi.org/10.17909/51hc-z182}.

Background correction was applied using the \texttt{Background2D} function from the \texttt{photutils.background} module. 
This method estimates the background by computing local statistics within a grid of boxes across the input image. 
A low-resolution background map is generated and then interpolated to produce a full-resolution version, which is subtracted from the original image to yield a 
background-corrected product.
The \texttt{astroalign} package was then used for the alignment of the NIRCam images. 
This routine detects bright sources in both the base and target images, matches them optimally, and computes a transformation to align the target image to the base. 
The algorithm relies solely on pixel coordinates and is independent of the initial image resolution. 
The output is an aligned image suitable for direct pixel-by-pixel comparison (\citealt{2020A&C....3200384B}). 
The FITS header of the aligned image is then updated to maintain accurate coordinate information. 
To assess the alignment quality or track proper motions, a script was developed to compute the average, minimum, median, and maximum distance differences between the two sets of stellar coordinates. 
These differences were provided in both pixels and arcseconds when the pixel scale was available. 
The aligned F212N image showed deviations of $<0.05\arcsec$ relative to the F460M image, indicating positional consistency across filters. 
This procedure was similarly applied to align the Subaru H$\alpha$ (see \S \ref{sec:ancillary}) and NIRCam F460M data.
 
\subsection{ALMA and Subaru Observations} \label{sec:ancillary}
Ancillary ALMA data were obtained from Cycle 6 observations of project 2018.1.01038.S. 
Band 6, 1.3 mm observations toward HH270VLA1 were conducted in the C-5 configuration with 43 to 50 operating antennas, and enabled an angular resolution of $\sim0.3 \arcsec$. 
The observations were carried out with 8 executions between Oct-02-2018 and Nov-23-2018. 
The monitored quasar J0510+1800 was used as the flux density and bandpass calibrator and the quasar J0532+0732 was used as the phase calibrator. 
The total time on source was $\sim$32 min. 
The dataset includes spectral line observations of the $J=2 \to 1$ transition of C$^{18}$O (219.560 GHz), $^{13}$CO (220.399 GHz), and $^{12}$CO (230.538 GHz). 
The data were calibrated by the pipeline integrated within \texttt{CASA 5.4.0-70} (\citealt{2022PASP..134k4501C}). 
The ALMA imaging pipeline was run manually on the HH270VLA1 calibrated data to find the spectral ranges that traced continuum emission using the \texttt{hif\_findcont} task. 
The continuum was subsequently subtracted from the spectral line visibilities using the pipeline tasks \texttt{hif\_uvcontfit} and \texttt{hif\_uvcontsub}. 
First, phase-only self-calibration was performed on the continuum data, and amplitude self-calibration was then performed, pre-applying the final phase-only solution interval and solving for both phase and amplitude solutions. 
The self-calibration solutions were then applied to the spectral line data. 
While, self-calibration had little impact on the C$^{18}$O and $^{13}$CO, it marginally increased the S/N of the $^{12}$CO data. 
Following continuum subtraction and self-calibration, final images of the spectral line data were made using \texttt{tclean}. 
The C$^{18}$O and $^{13}$CO data were imaged with 120, 0.25 km s$^{-1}$ channels, with 960$\times$960 0.048$\arcsec$ pixels, and a beam size of $0.29\arcsec \times 0.26\arcsec$. 
The $^{12}$CO data was imaged with 200, 1 km s$^{-1}$  channels, with 960$\times$960 0.045$\arcsec$ pixels, and a $0.28\arcsec \times 0.24\arcsec$ beam size. 
Rather than manually masking all the spectral line channels, the automated masking in \texttt{tclean} was used via the \texttt{usemask=`auto-multithresh'} option in \texttt{tclean} (\citealt{2020PASP..132b4505K}).

H$\alpha$ observations of the HH 270/110 system were taken with the Subaru Telescope on January 4, 2006 with the Subaru Prime Focus Camera (with 0.20 arcsec pixel$^{-1}$), which were published in \cite{2012AJ....143..106K}. The total time on source was 1 hr, and the FWHM of the point-spread function of the H$\alpha$ images is 1.1\arcsec. See \cite{2012AJ....143..106K} for details on the data reduction and image creation procedures.

\section{Results} \label{sec:Results}
Knot positions potentially associated with the HH270 outflow were identified in the NIRCam F460M and F212N images. 
These positions were refined using the \texttt{centroid\_sources} and \texttt{centroid\_com} routines from the \texttt{photutils.centroids} module. 
A script was developed to fit centroids in the corrected images at each of the knot coordinates from the NIRCam data, in order to find the coordinate corresponding to the peak emission of each potential knot. 
The script also calculated the average, minimum, median, and maximum positional differences between the initial and refined coordinates, which allowed us to assess that the refinement of the initial estimates was, in fact, minimal. 
To further enhance the visibility of compact structures in the images during the identification process, a Ricker wavelet of the form 
\begin{equation}
    f(x,\ y)=A\left[1-\frac{x^2 + y^2}{2\sigma^2}\right] \exp{\left[-\frac{x^2 + y^2}{2\sigma^2}\right]}
\end{equation}
where $A=1/\pi \sigma^4$ is the amplitude and $\sigma$ is the width, was applied in 2D convolution as a bandpass filter to smooth the data and suppress slowly-varying or constant structures, such as background gradients (\citealt{2015GeoJI.200..111W}).

\begin{figure}[htpb]
    \centering
    \includegraphics[width=0.38\linewidth]{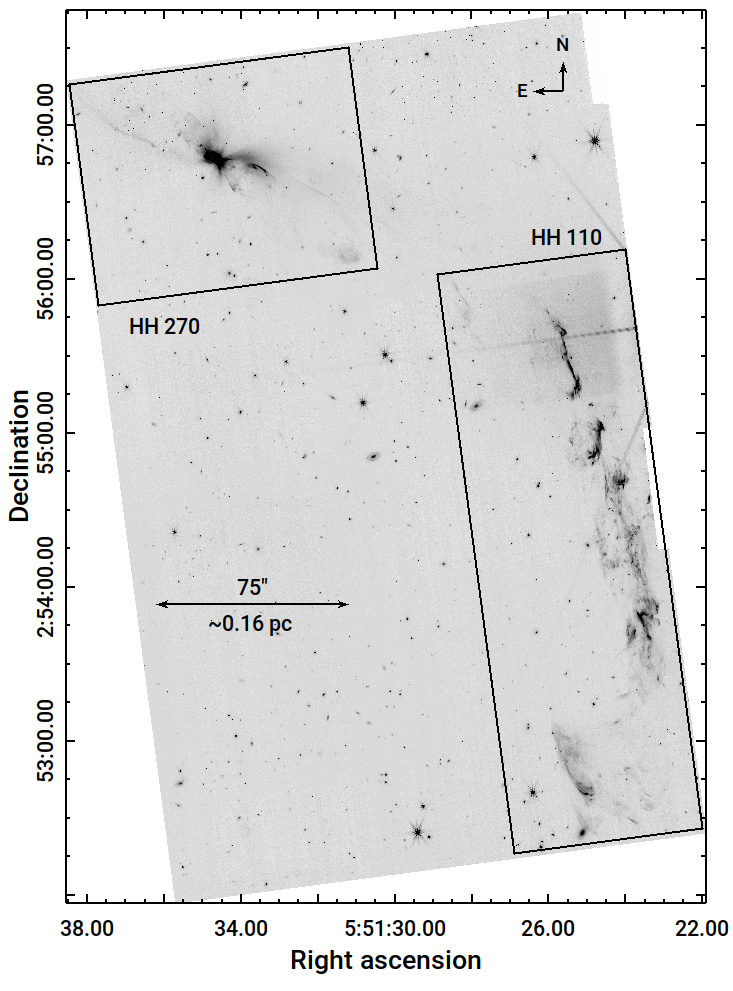}
    \includegraphics[width=0.61\linewidth]{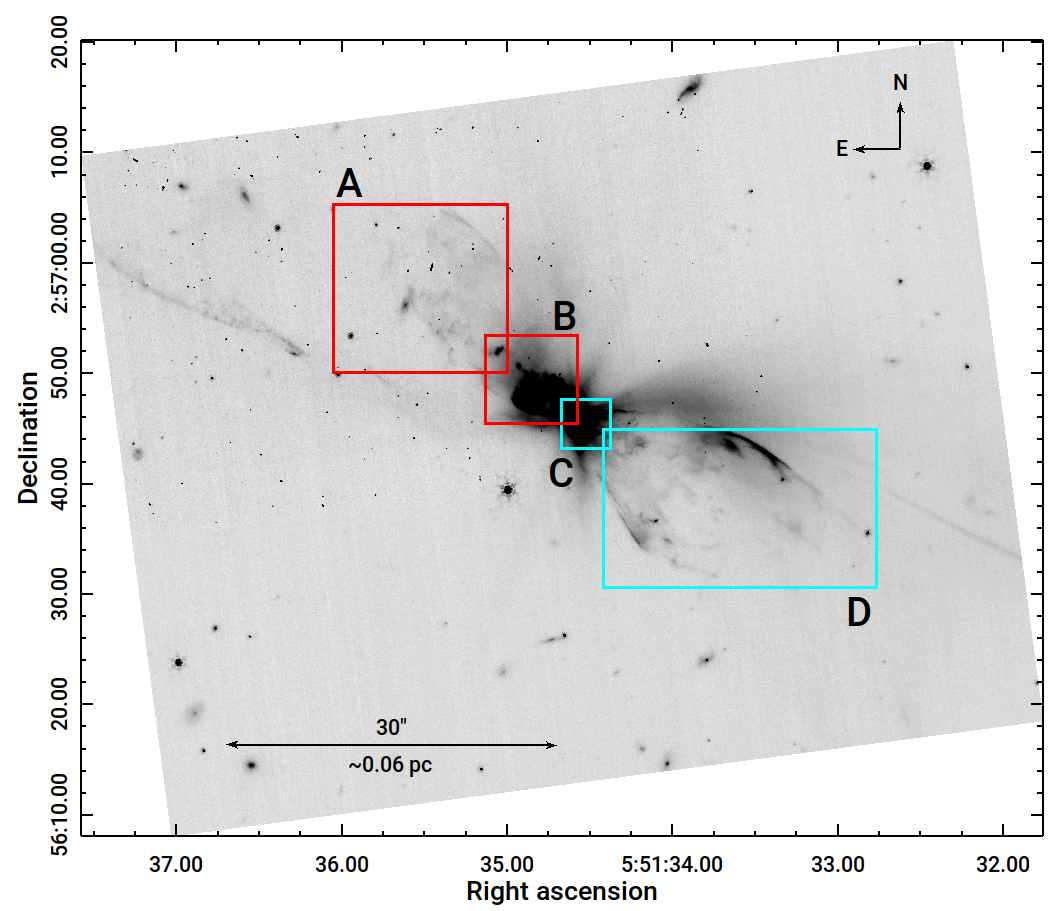}
    \caption{HH 270 observed with NIRCam F460M filter (4.60 $\mu$m), which is sensitive to H$_2$ and CO emission. Left panel shows the entire field, in which a highly detailed bipolar cavity and jet are visible in HH 270. The HH 110 shock fronts are also prominent in the image. To visualize the jet, the right panel shows the HH 270 field fragmented into four labeled subregions.}
    \label{fig: JWST F460M}
\end{figure}

\begin{figure}[htpb]
    \centering
    \includegraphics[width=0.49\linewidth]{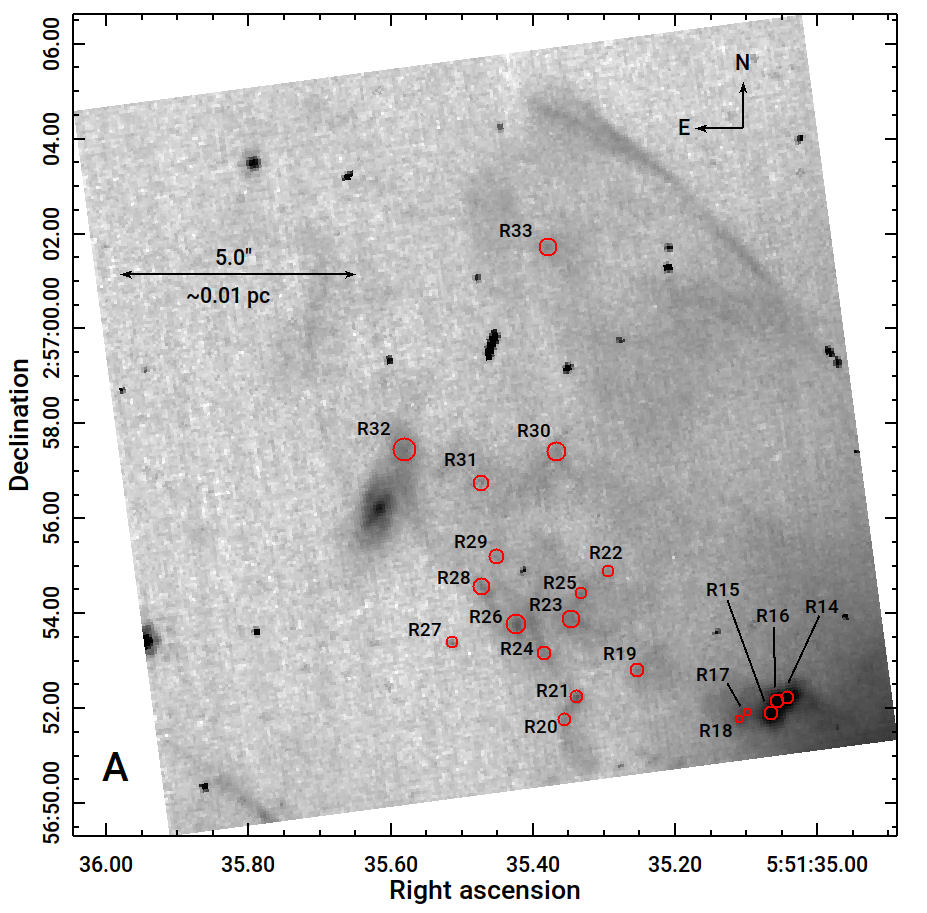}
    \includegraphics[width=0.49\linewidth]{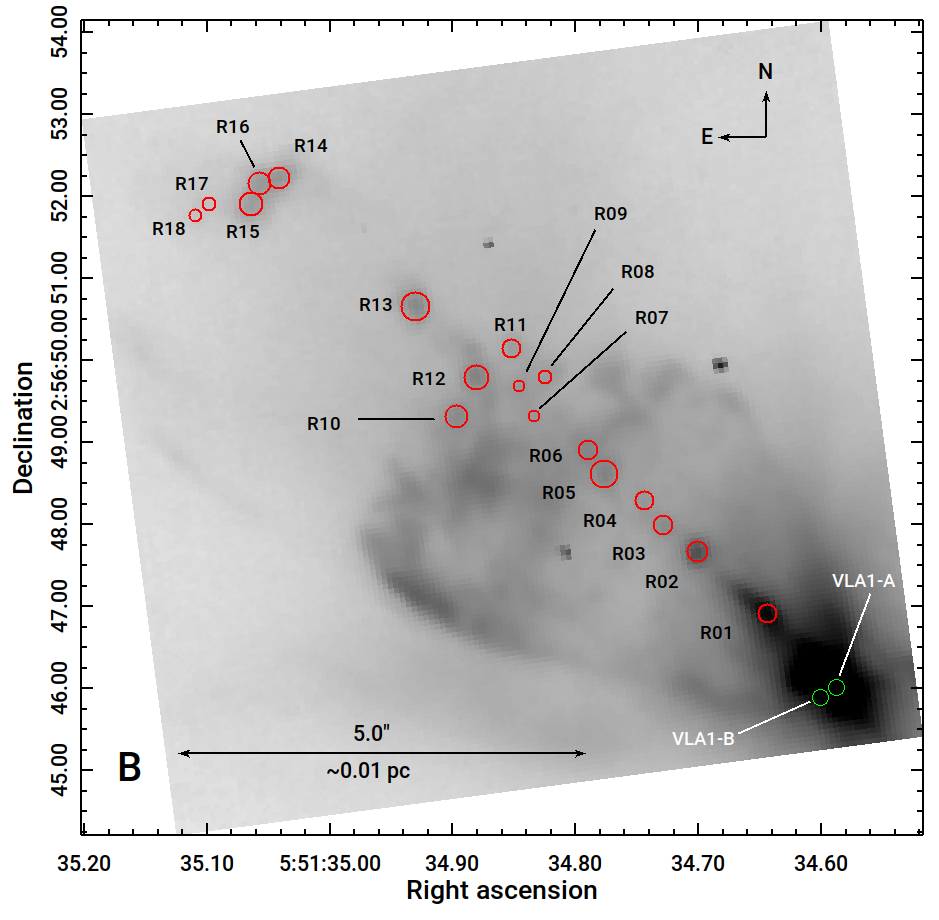}    
    \includegraphics[width=0.49\linewidth]{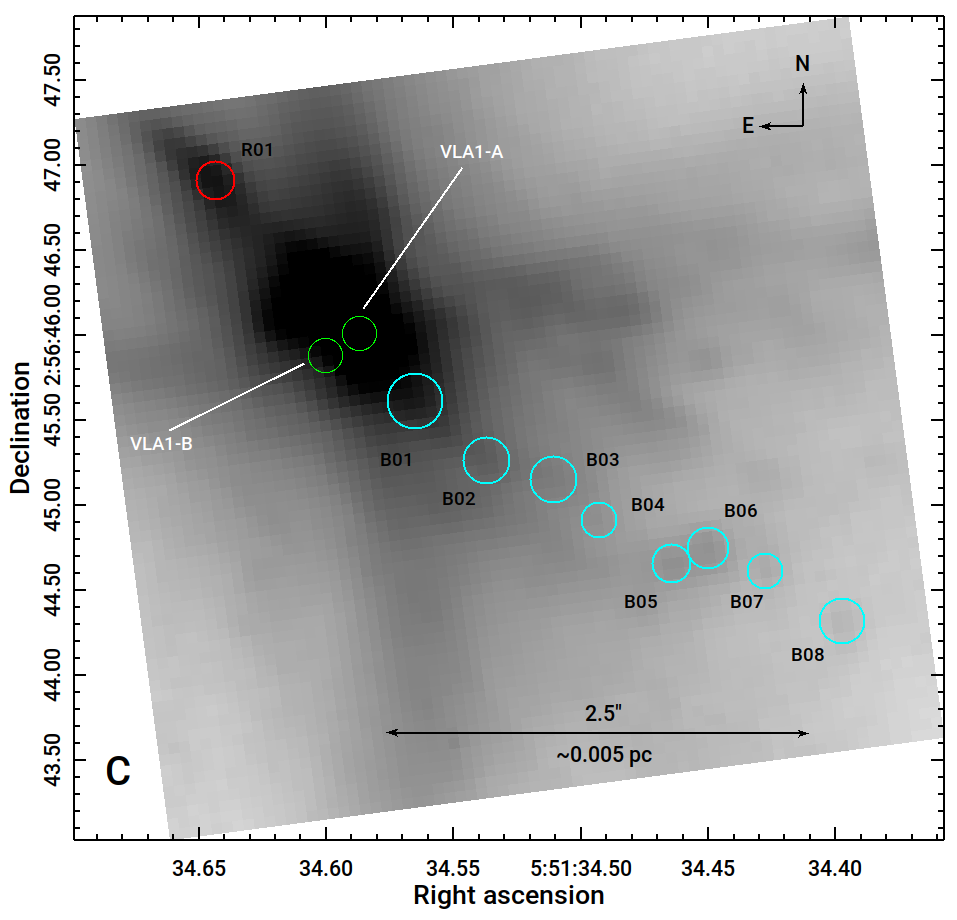}
    \includegraphics[width=0.49\linewidth]{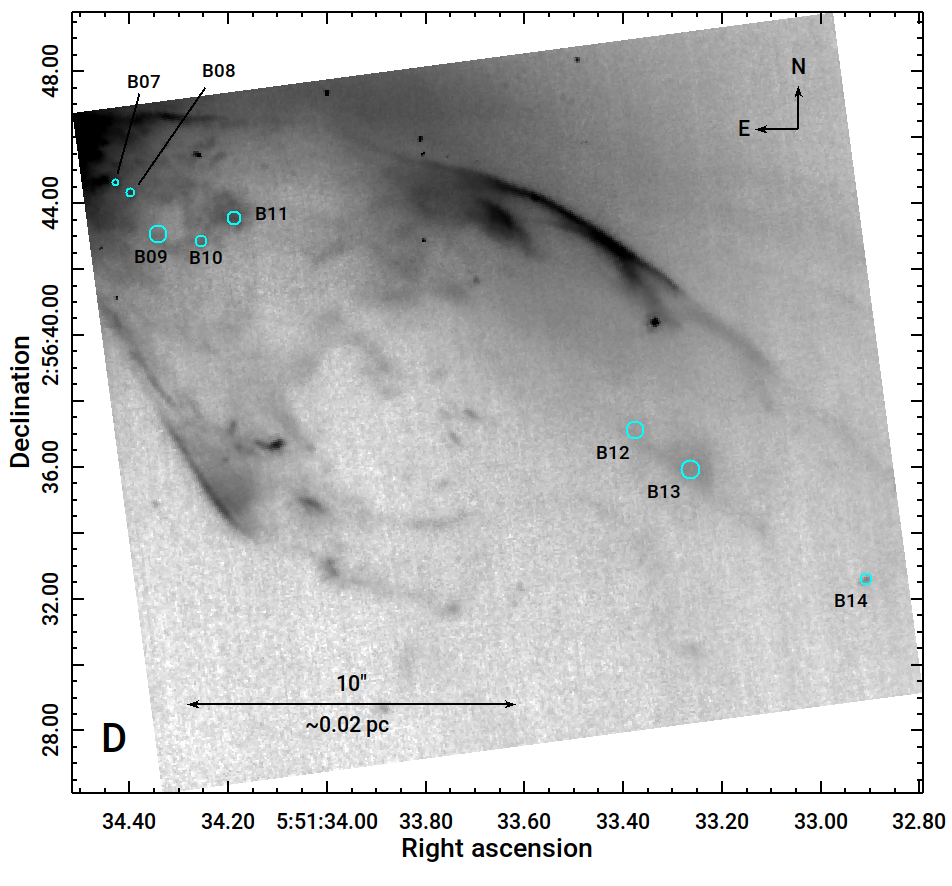}
    \caption{Panels A to D (as previously defined in Fig. \ref{fig: JWST F460M}) show color-coded regions highlighting the kinematic orientation of the HH 270 outflow: red denotes redshifted lobe (moving away from the observer), blue indicates the blueshifted lobe (moving toward the observer), and green represents the location of the central sources (VLA1-A and VLA1-B) identified in ALMA continuum data (\citealt{2022ApJ...925...39T}). The knot positions identified in the image are listed in Table \ref{tab: Jets Knots F460M}. The wavelet filtering was used only during the knot identification process and is not applied in the images shown in this article.} 
    \label{fig: JWST Boxes}
\end{figure}

Figure \ref{fig: JWST F460M} shows the HH 270/110 field as observed with the NIRCam F460M filter. 
To characterize the morphology of the HH 270 jet, four subregions were defined: two in the redshifted lobe (regions A and B) and two in the blueshifted lobe (regions C and D; see Figure \ref{fig: JWST Boxes}). 
Table \ref{tab: Jets Knots F460M} includes all detected knots with the F460M filter, while Table \ref{tab: Jets Knots F212N} reports knots exclusively found with the F212N filter. 
The naming convention assigns ``R'' for redshifted and ``B'' for blueshifted knots, with ascending numbers corresponding to increasing distance from the central source. 
Table \ref{tab: Jets Knots ALMA12CO} shows the redshifted knots identified through visual inspection in the ALMA $^{12}$CO emission, where most of them (58\%) can be found within a distance of $\leq10\arcsec$ from the central source. 
The knots in the ALMA $^{12}$CO moment maps were only identified manually, because the lower spatial resolution and extended morphology of the molecular emission made the use of the centroid fitting routines less reliable.

\begin{table}[htpb]
    \centering
    \caption{Knots associated with the HH 270 outflow identified in the F460M filter.}
    \label{tab: Jets Knots F460M}
    \begin{tabular}{l c c r r c}
        \toprule
        & & & \multicolumn{2}{c}{Distance from} & Corresponding \\
        Knots\tnote{1} & R.A. & Dec & \multicolumn{2}{c}{central source} & source in \\
        & (J2000) & (J2000) & (arcsec) & (au)\tnote{$\dagger$} & F212N filter \\
        \midrule
            R01* & 5:51:34.6432 & +2:56:46.911 & 1.22 & 524.56 & RH03 \\
            R02* & 5:51:34.7002 & +2:56:47.661 & 2.35 & 1008.63 & RH04 \\
            R03* & 5:51:34.7281 & +2:56:47.984 & 2.87 & 1233.36 & RH05\\
            R04* & 5:51:34.7439 & +2:56:48.290 & 3.25 & 1398.67 & RH06\\
            R05* & 5:51:34.7764 & +2:56:48.606 & 3.82 & 1642.74 & RH07 \\
            R06* & 5:51:34.7896 & +2:56:48.895 & 4.16 & 1790.59 & RH08 \\
            R07* & 5:51:34.8334 & +2:56:49.320 & 4.93 & 2120.44 & RH09 \\
            R08 & 5:51:34.8248 & +2:56:49.788 & 5.17 & 2225.51 & ---\\
            R09 & 5:51:34.8451 & +2:56:49.683 & 5.31 & 2283.11 & ---\\
            R10* & 5:51:34.8967 & +2:56:49.314 & 5.66 & 2432.26 & RH11 \\
            R11 & 5:51:34.8517 & +2:56:50.142 & 5.71 & 2454.79 & --- \\
            R12* & 5:51:34.8808 & +2:56:49.788 & 5.77 & 2481.62 & RH12 \\
            R13* & 5:51:34.9302 & +2:56:50.654 & 6.90 & 2967.84 & RH13\\
            R14* & 5:51:35.0412 & +2:56:52.221 & 9.20 & 3954.83 & RH14\\
            R15* & 5:51:35.0640 & +2:56:51.896 & 9.23 & 3970.51 & RH15 \\
            R16* & 5:51:35.0566 & +2:56:52.144 & 9.30 & 4001.41 & RH16\\
            R17 & 5:51:35.0979 & +2:56:51.907 & 9.62 & 4139.68 & ---\\
            R18* & 5:51:35.1093 & +2:56:51.760 & 9.67 & 4159.25 & RH17\\
            R19 & 5:51:35.2536 & +2:56:52.799 & 12.03 & 5174.79 & ---\\
            R20 & 5:51:35.3551 & +2:56:51.745 & 12.80 & 5504.85 & ---\\
            R21 & 5:51:35.3386 & +2:56:52.247 & 12.82 & 5513.02 & ---\\
            R22 & 5:51:35.2937 & +2:56:54.878 & 13.78 & 5925.70 & ---\\
            R23 & 5:51:35.3457 & +2:56:53.877 & 13.78 & 5926.73 & ---\\
            R24 & 5:51:35.3840 & +2:56:53.157 & 13.87 & 5963.38 & ---\\
            R25 & 5:51:35.3311 & +2:56:54.407 & 13.92 & 5985.87 & ---\\
            R26 & 5:51:35.4224 & +2:56:53.760 & 14.67 & 6310.30 & ---\\
            R27 & 5:51:35.5126 & +2:56:53.375 & 15.65 & 6728.91 & ---\\
            R28 & 5:51:35.4713 & +2:56:54.555 & 15.72 & 6760.14 & ---\\
            R29 & 5:51:35.4503 & +2:56:55.193 & 15.82 & 6804.05 & ---\\
            R30 & 5:51:35.3666 & +2:56:57.404 & 16.29 & 7007.19 & ---\\
            R31 & 5:51:35.4720 & +2:56:56.740 & 17.02 & 7320.82 & ---\\
            R32 & 5:51:35.5797 & +2:56:57.444 & 18.72 & 8051.90 & ---\\
            R33 & 5:51:35.3792 & +2:57:01.722 & 19.68 & 8466.09 & --- \\
            \hline
            B01* & 5:51:34.5649 & +2:56:45.613 & 0.54 & 233.11 & BH01\\
            B02* & 5:51:34.5367 & +2:56:45.261 & 1.09 & 469.53 & BH02\\
            B03 & 5:51:34.5105 & +2:56:45.151 & 1.48 & 634.48 & ---\\
            B04 & 5:51:34.4926 & +2:56:44.910 & 1.83 & 787.89 & ---\\
            B05* & 5:51:34.4643 & +2:56:44.656 & 2.33 & 1000.15 & BH06\\
            B06* & 5:51:34.4500 & +2:56:44.749 & 2.46 & 1058.02 & BH07\\
            B07	 & 5:51:34.4275 & +2:56:44.611 & 2.82 & 1213.69 & ---\\
            B08* & 5:51:34.3975 & +2:56:44.317 & 3.36 & 1443.95 & BH08 \\
            B09 & 5:51:34.3412 & +2:56:43.039 & 4.77 & 2050.50 & ---\\
            B10 & 5:51:34.2535 & +2:56:42.844 & 5.96 & 2564.67 & ---\\
            B11* & 5:51:34.1887 & +2:56:43.533 & 6.53 & 2806.84 & BH09\\
            B12	 & 5:51:33.3747 & +2:56:37.097 & 20.29 & 8726.30 & ---\\
            B13	 & 5:51:33.2635 & +2:56:35.900 & 22.31 & 9596.82 & ---\\
            B14* & 5:51:32.9067 & +2:56:32.602 & 28.58 & 12290.11 & BH10\\
            \bottomrule
    \end{tabular}
     \begin{minipage}{\textwidth}
     \footnotesize
     \begin{itemize}
         \item[1] Labels starting with ``R'' correspond to knots in the redshifted lobe, while ``B'' indicates knots in the blueshifted lobe.
        \item[*] Asterisks denote knots with corresponding detections in the F212N filter.
        \item[$\dagger$] To calculate the distance in Astronomical Units (au) we assume a distance of 430.1 pc to HH270VLA1  (\citealt{2022ApJ...925...39T}).
     \end{itemize}
     \end{minipage}
\end{table}

\begin{table}[htpb]
        \centering
        \caption{Knots exclusively identified in the HH 270 outflow using the F212N filter.}
        \label{tab: Jets Knots F212N}
        \begin{tabular}{l c c r r}
            \toprule
        & & & \multicolumn{2}{c}{Distance from} \\
        Knots & R.A. & Dec & \multicolumn{2}{c}{central source} \\
        & (J2000) & (J2000) & (arcsec) & (au) \\
        \midrule
            RH01 & 5:51:34.6406 & +2:56:46.528 & 0.92 & 393.66 \\
            RH02 & 5:51:34.6548 & +2:56:46.759 & 1.23 & 527.80 \\
            RH10 & 5:51:34.8310 & +2:56:49.635 & 5.13 & 2204.62 \\
            \hline
            BH03 & 5:51:34.5588 & +2:56:44.889 & 1.18 & 506.22 \\
            BH04 & 5:51:34.5240 & +2:56:44.860 & 1.50 & 646.75 \\
            BH05 & 5:51:34.4954 & +2:56:44.531 & 2.04 & 877.12 \\
            \bottomrule
        \end{tabular}
        \begin{minipage}{9cm}
            \vspace{0.2cm}
            \footnotesize
            Note: Jet knots exclusively identified in the F212N filter. The ``R'' and ``B'' labels follow the same convention as in Table \ref{tab: Jets Knots F460M}, corresponding to redshifted and blueshifted lobes, respectively.
        \end{minipage}
    \end{table}

\begin{table}[htpb]
        \centering
        \caption{Knots associated with the redshifted lobe in ALMA $^{12}$CO $(J=2\to 1)$ emission.}
        \label{tab: Jets Knots ALMA12CO}
        \begin{tabular}{l c c r r}
            \toprule
        & & & \multicolumn{2}{c}{Distance from} \\
        Knots & R.A. & Dec & \multicolumn{2}{c}{central source} \\
        & (J2000) & (J2000) & (arcsec) & (au) \\
        \midrule
            RCO01 & 5:51:34.6308 & +2:56:46.563 & 0.83 & 358.24 \\
            RCO02 & 5:51:34.6460 & +2:56:46.749 & 1.12 & 483.75 \\
            RCO03 & 5:51:34.6721 & +2:56:47.005 & 1.58 & 681.24 \\
            RCO04 & 5:51:34.6911 & +2:56:47.360 & 2.03 & 874.97 \\
            RCO05 & 5:51:34.7192 & +2:56:47.655 & 2.54 & 1093.93 \\
            RCO06 & 5:51:34.7404 & +2:56:48.021 & 3.03 & 1301.12 \\
            RCO07 & 5:51:34.7646 & +2:56:48.483 & 3.61 & 1551.27 \\
            RCO08 & 5:51:34.8309 & +2:56:49.558 & 5.07 & 2180.47 \\
            RCO09 & 5:51:34.8736 & +2:56:49.534 & 5.52 & 2375.09 \\
            RCO10 & 5:51:34.9386 & +2:56:50.571 & 6.94 & 2983.69 \\
            RCO11 & 5:51:34.9917 & +2:56:51.309 & 8.02 & 3450.25 \\
            RCO12 & 5:51:34.9971 & +2:56:51.520 & 8.22 & 3537.17 \\
            RCO13 & 5:51:35.0800 & +2:56:51.796 & 9.35 & 4019.64 \\
            RCO14 & 5:51:35.1155 & +2:56:51.444 & 9.56 & 4111.55 \\
            RCO15 & 5:51:35.1200 & +2:56:51.974 & 9.93 & 4269.96 \\
            RCO16 & 5:51:35.1409 & +2:56:51.750 & 10.05 & 4321.13 \\
            RCO17 & 5:51:35.2070 & +2:56:51.942 & 10.97 & 4719.92 \\
            RCO18 & 5:51:35.1908 & +2:56:52.533 & 11.11 & 4778.97 \\
            RCO19 & 5:51:35.2207 & +2:56:52.533 & 11.48 & 4935.53 \\
            RCO20 & 5:51:35.2695 & +2:56:53.194 & 12.45 & 5356.12 \\
            RCO21 & 5:51:35.2714 & +2:56:54.038 & 12.99 & 5584.85 \\
            RCO22 & 5:51:35.2979 & +2:56:53.935 & 13.24 & 5692.76 \\
            RCO23 & 5:51:35.2842 & +2:56:54.401 & 13.36 & 5747.34 \\
            RCO24 & 5:51:35.3219 & +2:56:54.033 & 13.58 & 5841.75 \\
            RCO25 & 5:51:35.3479 & +2:56:54.266 & 14.03 & 6036.24 \\
            RCO26 & 5:51:35.3884 & +2:56:54.672 & 14.76 & 6349.65 \\
            \bottomrule
        \end{tabular}
        \begin{minipage}{9cm}
            \vspace{0.2cm}
            \footnotesize
            Note: As in the NIRCam measurements, we used the distance of 430.1 pc, which corresponds to HH270VLA1 (see \citealt{2022ApJ...925...39T}).
        \end{minipage}
    \end{table}

ALMA continuum observations at 0.87 mm confirmed two central components, VLA1-A and VLA1-B, located at $\alpha = 5:51:34.587$, $\delta=+2:56:46.010$ and $\alpha=5:51:34.600$, $\delta=+2:56:45.880$, respectively (\citealt{2022ApJ...925...39T}). 
These locations were used as reference points to anchor the jet's position angle and determine knot distribution relative to the source.

\begin{figure}[htpb]
    \centering
    \includegraphics[width=\linewidth]{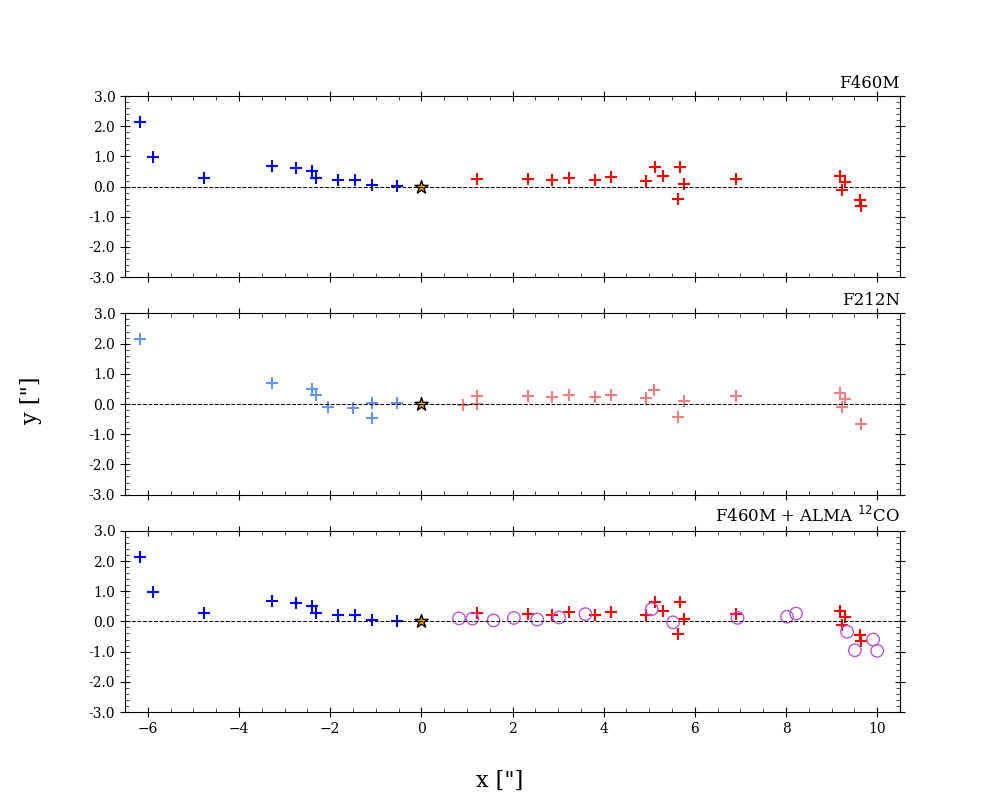}
    \caption{Positions of inner HH 270 jet knots relative to the central source (set at (0, 0)) on both NIRCam filters and ALMA $^{12}$CO emission. Red and blue crosses represent redshifted and blueshifted knots, respectively. Purple circles represent ALMA's redshifted knots. This visualization highlights deviations from linearity, asymmetries between lobes, and structural features that may be linked to precession or jet-environment interactions.}
    \label{fig: Nearby knots from Center}
\end{figure}

To better visualize the morphology and symmetry of the HH 270 jet, we plotted the spatial distribution of the identified knots relative to the central source, defined as the origin (0, 0). 
In this coordinate system, we choose to orient the redshifted knots towards the positive \textit{x-y} quadrant. 
Then, to align the outflow axis along the \textit{x} coordinate, all knots' positions were rotated clockwise by $\sim$43 degrees (see Fig \ref{fig: Nearby knots from Center}). 
This method helps reveal the degree of collimation, possible signs of precession of the jet, and asymmetries in the flow pattern of HH 270. 

At distances close to the central source ($\leq$ 5\arcsec), the redshifted and blueshifted knots seem to be aligned along a common axis (see top panel of Fig. \ref{fig: Nearby knots from Center}), indicating that the inner jet remains highly collimated. 
And when overlaid, the knots identified in the $^{12}$CO data from ALMA (bottom panel in Fig. \ref{fig: Nearby knots from Center}), although only seen in the redshifted region, align very well with the NIR knots from the NIRCam data. 
This implies that, at longer distances from the central system, the redshifted $^{12}$CO jet can be seen as an extension of the NIR jet. 

\begin{figure}[htpb]
    \centering
    \includegraphics[width=\linewidth]{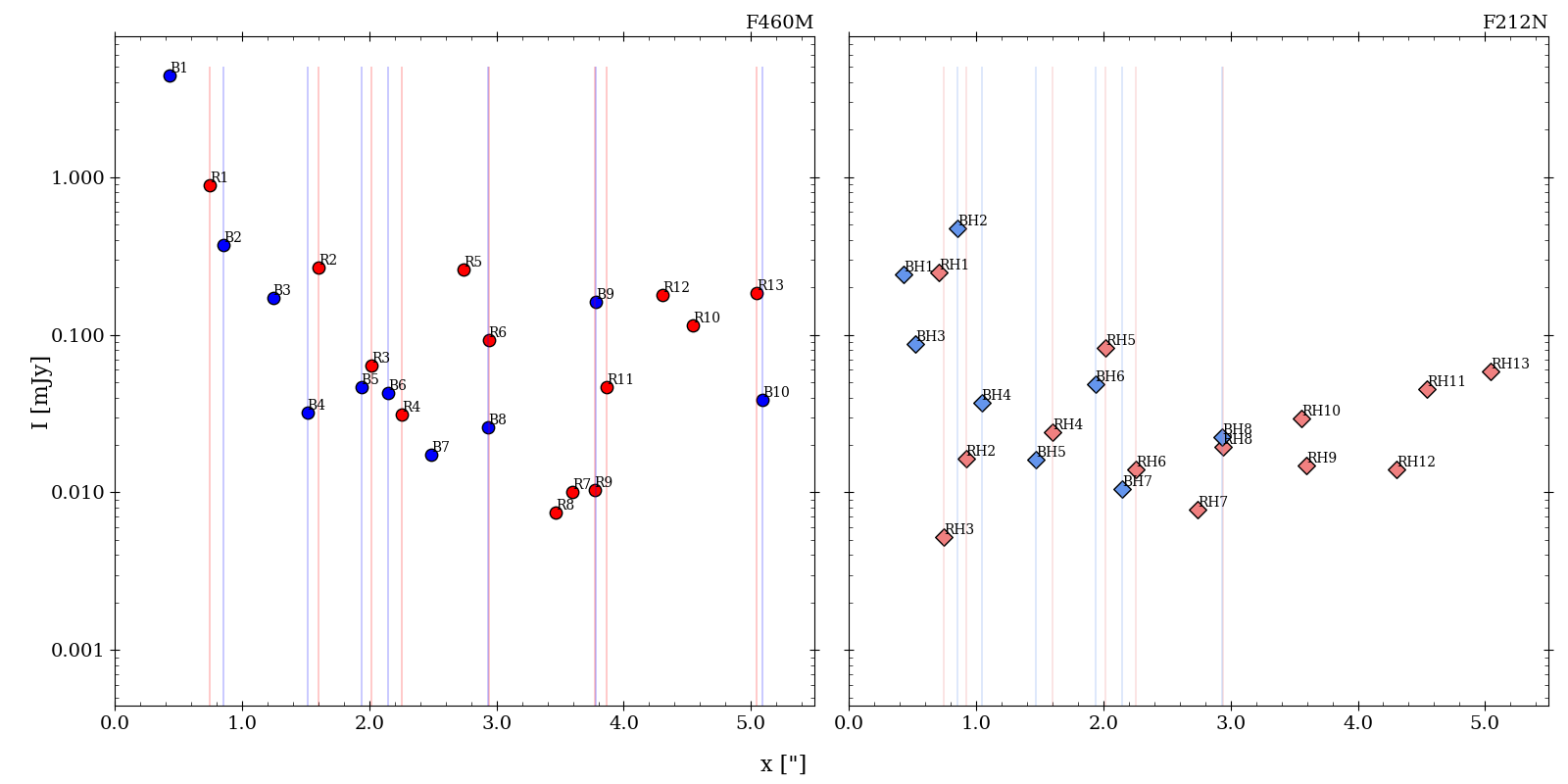}
    \caption{Background-subtracted fluxes of identified HH 270 jet knots as a function of distance \textit{x} from the central source. The left panel shows the subset of knots located close to the central source in F460M emission, while the right panel shows the identified knots in F212N emission.}
    \label{fig: Knots Fluxes}
\end{figure}

The flux of each identified knot was measured using a circular region which encompasses the full extent of each knot. 
Background emission was subtracted using a nearby circular region of an area identical to the size of each particular knot, located at $\sim$ 5\arcsec away from it, and selected to avoid contamination from adjacent jet features.
These background-subtracted fluxes were then plotted as a function of the distance \textit{x} from the central source for both lobes and in both NIRCam filters (see Fig. \ref{fig: Knots Fluxes}). 
This method, like that employed by \cite{2020RMxAA..56...29N} for the HH 212 jet, enables an analysis of the knot brightness evolution along the jet and allows for the identification of asymmetries or episodic variations in the outflow. 

The vertical lines in Figure \ref{fig: Knots Fluxes} highlight the possible occurrence of redshifted/blueshifted knot pairs with closely matched positions. 
In the F460M and F212N emission, most (70\% and 75\%, respectively) of the blueshifted knots have a redshifted counterpart within a distance of 0.13\arcsec, and the majority of the knot pairs occur $<3\arcsec$ from the central source. In the F460M filter, most (71\%) of the redshifted knots are brighter than the blueshifted knots, and there seems to be no common factor in the ratios between the redshifted and blueshifted intensities. 
In the F212N emission, about half of the redshifted knots are brighter than the blueshifted knots, and the ratio of the redshifted/blueshifted knot intensities seems to decrease with increasing knot distance from the central source.

\begin{figure}[htpb]
    \centering
    \includegraphics[width=\linewidth]{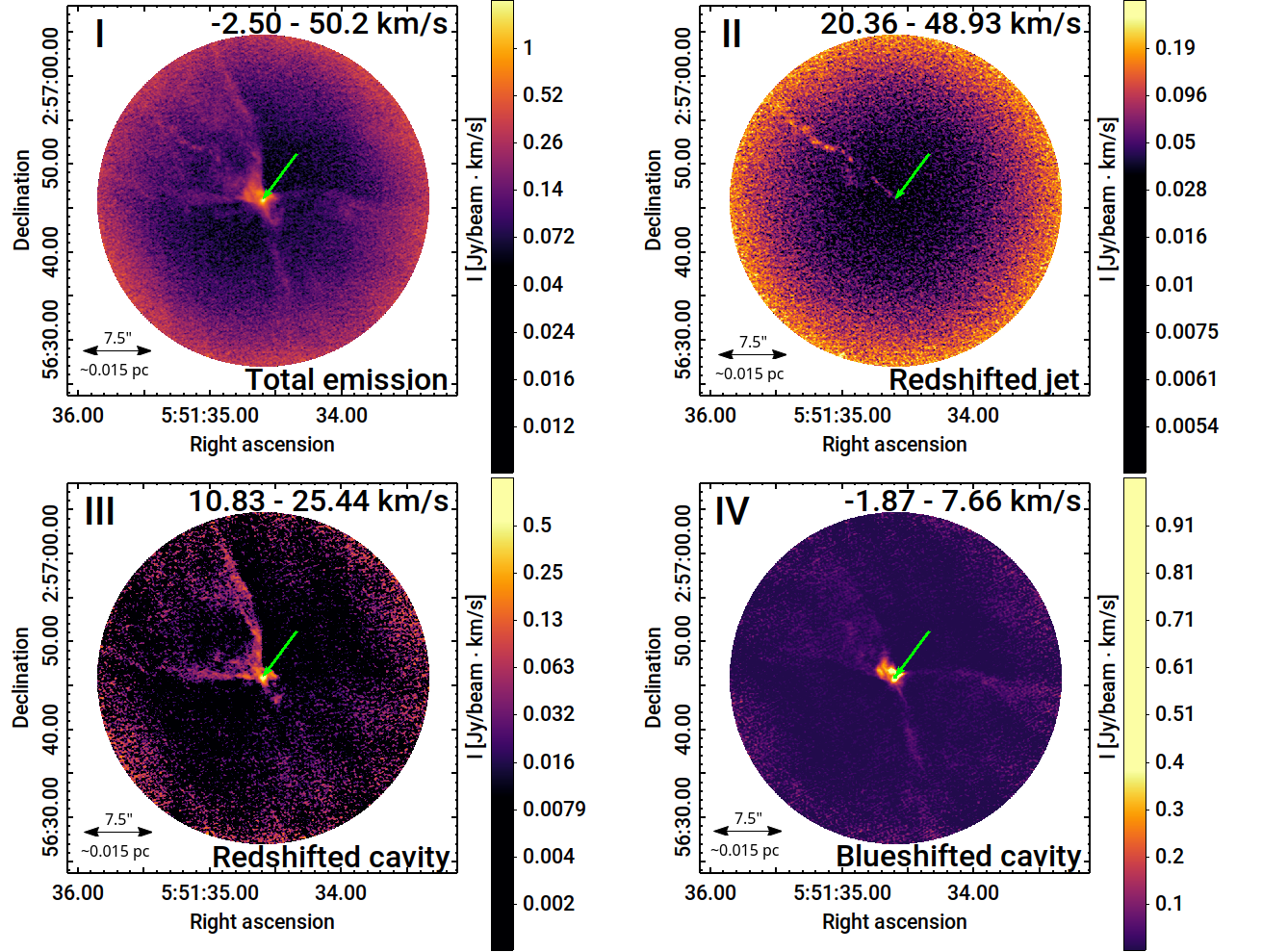}
    \caption{Moment-0 maps of $^{12}$CO ($J=2 \to 1$) emission in HH 270 across multiple velocity intervals. Panel I shows overall emission, Panel II isolates the redshifted jet, Panel III highlights the redshifted cavity structure, and Panel IV shows the blueshifted lobe, where fainter emissions reveals a partial cavity structure with no visible jet. The green arrow indicates the position of HH270VLA1.}
    \label{fig: ALMA 12CO}
\end{figure}

ALMA moment-0 (integrated intensity) maps were generated for the $^{12}$CO, $^{13}$CO and C$^{18}$O lines, with their corresponding ($J=2 \to 1$) transitions using the Cube Analysis and Rendering Tool for Astronomy (CARTA). 
These CO low-$J$ transitions effectively trace both high-velocity jets and broader low-velocity molecular outflows from YSOs. 
The $^{12}$CO emission is shown in Figure \ref{fig: ALMA 12CO}, subdivided into four panels: (I) total emission (-2.50 $\to$ 50.2 km s$^{-1}$), (II) redshifted jet (20.36 $\to$ 48.93 km s$^{-1}$), (III) redhisfted cavity (10.83 $\to$ 25.44 km s$^{-1}$) , and (IV) blueshifted cavity (-1.87 $\to$ 7.66 km s$^{-1}$). 
A logarithmic scale was applied to enhance faint structures. 
In panel (II), the outflow extends linearly from $\sim 500$ au to $\sim4000$ au from the central source, curving beyond that, a trend consistent with jet precession or deflection. 
In contrast, panel (III) traces intermediate-velocity gas associated with the cavity walls, and panel (IV) shows faint and difuse low-velocity emission coming from the cavity, both without a discernible jet.

\begin{figure}[htpb]
    \centering
    \includegraphics[width=\linewidth]{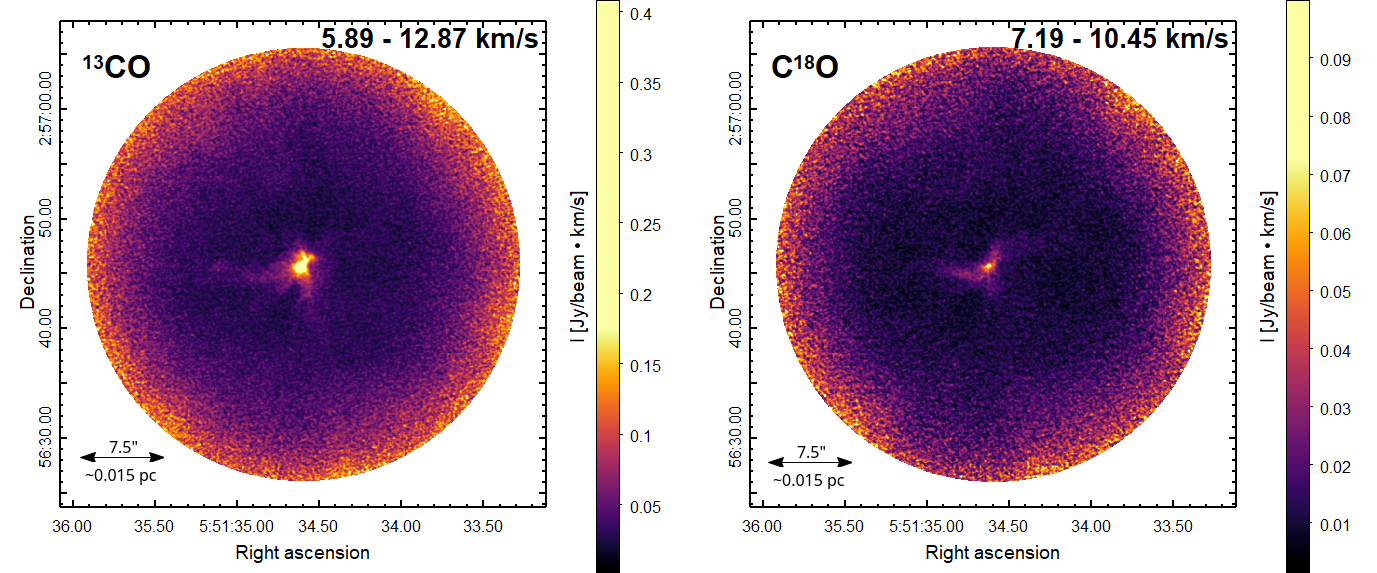}
    \caption{Moment-0 maps of $^{13}$CO and C$^{18}$O ($J=2 \rightarrow 1$) emissions on the HH 270 central source. Emission is concentrated near the central source, extending weakly into one side of the cavity in both redshifted and blueshifted directions. Unlike $^{12}$CO emissions, no jet is observed in these isotopologues, indicating a focus on denser, slower-moving gas around the core region.}
    \label{fig: ALMA 13CO and C18O}
\end{figure}

The $^{13}$CO and C$^{18}$O emission maps (Fig. \ref{fig: ALMA 13CO and C18O}) reveal compact emission near the central source, extending slightly into the cavity. 
These tracers highlight dense, slow-moving gas likely associated with the envelope or cavity walls, rather than the high-velocity jet. 
The molecular cloud velocity near HH 270 is estimated at $\sim$8-9 km s$^{-1}$, based on the centroid velocities of these lines.

\begin{figure}[htpb]
    \centering
    \includegraphics[width=0.75\linewidth]{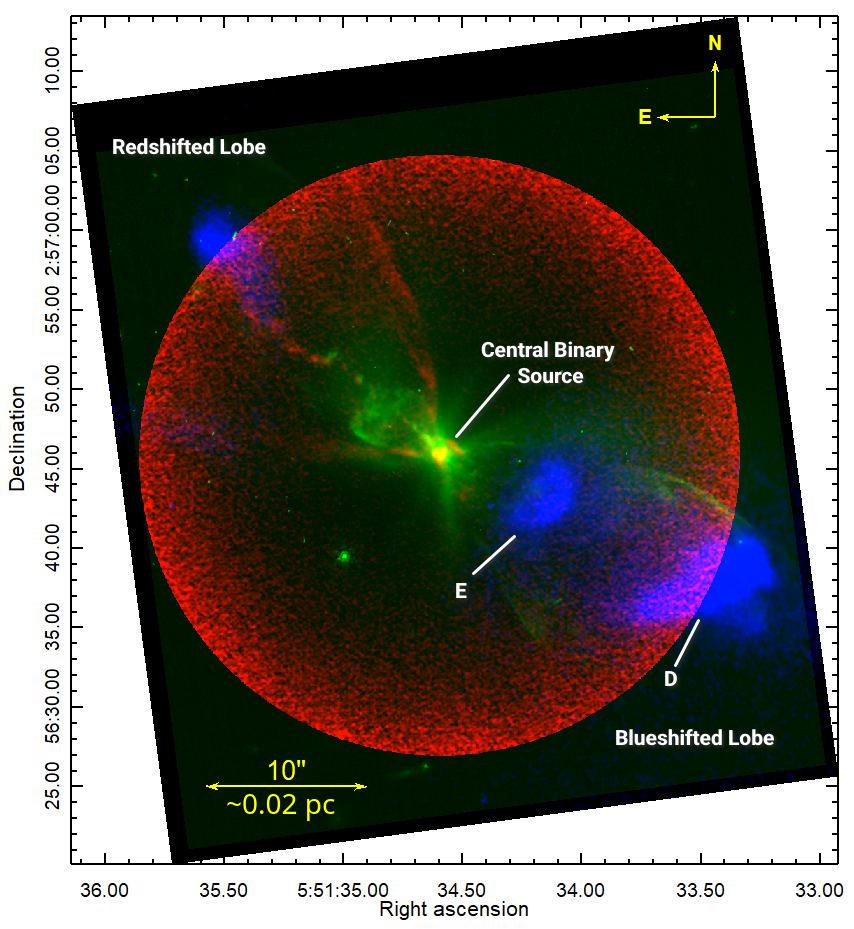}
    \caption{Comparative analysis of ALMA (1.3 mm, red), NIRCam F460M (4.60 $\mu$m, green), and Subaru H$\alpha$ (660 nm, blue) emissions, illustrating the multi-wavelength structure of HH 270's outflows. The alignment across wavelengths provides a comprehensive view of the jet's distribution and highlights differences in observed structures at optical, infrared, and radio wavelengths.}
    \label{fig: Comparison Image}
\end{figure}

A multi-wavelength comparison (Fig. \ref{fig: Comparison Image}) combines ALMA's redshifted $^{12}$CO emission (red), NIRCam F460M (green), and Subaru's H$\alpha$ data (blue). 
The alignment across wavelengths shows a strong morphological continuity in the redshifted lobe, where jet emission extends consistently from radio through IR and optical bands, which may indicate that the same underlying outflow structure is being observed through the different tracers: optical emission from ionized shocks (H$\alpha$), IR emission from molecular hydrogen shocks, and millimeter emission from entrained molecular gas.

\section{Discussion} \label{sec:Discussion}
The NIRCam images reveal a remarkable bipolar cavity structure in HH 270, with redshifted (upper) and blueshifted (lower) components clearly visible for the first time at such high spatial resolution. 
The F460M and F212N data allowed for the identification of numerous new jet knots, with the redshifted component showing a well-collimated jet extending outward from the central source (see Fig. \ref{fig: JWST Boxes}). 
In the redshifted lobe, the jet trajectory remains linear from knots R01 through R13 (from 1.22$\arcsec$ to 6.90$\arcsec$ NE from the central source). 
From the R14-R16 knots ($\sim9.20 \arcsec$ NE from the central source) and beyond this, the jet exhibits a slight curvature to the east and then moves to the north, forming an ``L''-shaped structure (see Fig. \ref{fig: JWST Boxes}, Panel B). 
A similar pattern is observed in the ALMA $^{12}$CO ($J=2 \to 1$)  emission between $\sim$20 and $\sim$49 km s$^{-1}$ (Fig. \ref{fig: ALMA 12CO}). 
This morphology supports two possible interpretations: (1) the central source is precessing, resulting in the observed jet curvature, which is a scenario consistent with models and observations of binary-driven precession in other HH objects like HH 30 (\citealt{2007AJ....133.2799A}, \citealt{2012AJ....144...61E}), HH 111 (\citealt{1997AJ....114..757R}, \citealt{2002ApJ...568..733M}) and HH 212 (\citealt{2020RMxAA..56...29N}); or (2) the jet is being deflected by an interaction with dense ambient gas in the surrounding molecular cloud.

In contrast to the features in the redshifted lobe, the blueshifted lobe appears fainter in the NIRCam images, suggesting a higher extinction towards the inner region close to the central source due to the dense molecular cloud environment (\citealt{2011A&A...527A..41S}). 
This faintness is consistent with earlier extinction studies that identified dense material along this side of the outflow. 
Also, disk extinction models like the ones presented in \cite{2024ApJ...966..225A} show that the extinction in this type of region could be as high as A$_{\lambda} >10$ mag at the wavelengths observed with NIRcam. 
The analysis carried out by \cite{2010ApJ...712.1010T} and \cite{2011ApJ...740...45T} using Spitzer mapping and N$_2$H$^+$ data traced the structure of the core surrounding HH270VLA1 and they found that the densest regions of the envelope lie preferentially along the blueshifted lobe. 
This is consistent with the attenuation seen in the NIRCam data (Fig. 2C and 2D) toward the blueshifted lobe. 
Furthermore, as seen in Figure 7, within $\sim5 \arcsec$ from the central source, the H$\alpha$ emission is not detected, likely due to extinction by a dense protostellar disk. 
Overall, the H$\alpha$ emission does not appear to be symmetric farther away from the disk, suggesting that the jet collimation process may be asymmetric even before crossing the disk boundary. 
Therefore, although extinction at optical wavelengths plays a role in what is detected, the outflow asymmetry in the infrared is likely due to the collimation process and the density distribution of the surrounding environment close to the driving source.

The first knot in the blueshifted lobe (B01) lies at $\sim$230 au from the central source. 
This proximity is particularly significant, as previous observations lacked the spatial resolution necessary to identify knots so close to the central source, which is an achievement made possible through JWST imaging capabilities for this HH object. 
The knots close to the source (B01 to B05) are approximately aligned along an axis with a position angle of $\sim$227 degrees, while beyond B06 the flow curves, similar to the redshifted lobe, consistent with the precession hypothesis (see Fig. 2, Panel C).
However, because not all knots are visible in both lobes, likely due to the higher extinction toward the blueshifted side, there is no strict one-to-one correspondence between redshifted and blueshifted knots. 
This is consistent with the interpretation that one lobe of the jet extends into the molecular cloud. 
Still, within $\leq5\arcsec$ from the central source, the knots in both directions remain approximately aligned along a common axis. 
This behavior is illustrated in Figure \ref{fig: Nearby knots from Center}, noting that the knots positions shown in the figure were rotated as described in 
\S \ref{sec:Results}.

Figure \ref{fig: Knots Fluxes} presents, for each filter, the background-subtracted flux of each knot as a function of its projected distance from the central source, for both redshifted and blueshifted lobe. 
Vertical lines indicate potential counterparts between blueshifted and redshifted knots in opposite lobes, based on their similar distance from the central source ($<0.13 \arcsec$ difference in distance between pairs). 
The presence of 8 such pairs in F460M and 7 in F212N emission supports the view that we are tracing genuine jet knots rather than unrelated features such as background emission.

The lack of perfect positional symmetry between possible knot pairs does not preclude physical correspondence. 
Such offset may arise from projection effect due to inclination differences between the two lobes, or from non-simultaneous ejection events. 
If a representative jet velocity of $\sim$200 km s$^{-1}$ is assumed (a typical value of HH flows; e.g. \citealt{2001ARA&A..39..403R}, \citealt{2010MNRAS.406.2193L}), the measured positional mismatches in HH 270 knots correspond to an ejection time difference on the order of a few to tens of years. 
This timescale is consistent with episodic variability observed in other protostellar jets and suggests that the identified pairs could still originate from the same underlying variability mechanism in the central source 
(\citealt{1998AJ....116.2943R}).

High-resolution ALMA continuum observations at 0.87 mm confirm the presence of a protobinary system (VLA1-A and VLA1-B; \citealt{2022ApJ...925...39T}), which challenges earlier single-source models based on VLA data (\citealt{1998RMxAA..34...69R}). 
It may be possible that the observed jet arises from combined o alternating activity from both sources, but the exact contribution of each protostar to the outflow remains uncertain given the current data.

ALMA's $^{12}$CO, $^{13}$CO, and C$^{18}$O ($J=2 \to 1$) observations primarily trace molecular gas that is being pushed by the protostellar wind, whereas the IR knots come from high-speed, shock-excited material along the jet's path. 
So even though these two components appear spatially related in both ALMA and JWST data, they may not arise from the same physical process. 
The $^{12}$CO line (Fig. \ref{fig: ALMA 12CO}, Panels II and III) reveals a redshifted outflow between 10.83 and 48.93 km s$^{-1}$. 
Within this range, a highly-detailed cavity and linear jet structure are visible at low velocities, transitioning into a curved, possibly precessing or deflected flow at higher velocities. 
This is consistent with the deflection observed in NIRCam images (Fig. \ref{fig: JWST Boxes}, Panel A). 
In the blueshifted range (Fig. \ref{fig: ALMA 12CO}, Panel IV), the emission is faint, with only a diffuse cavity present, lacking a prominent jet feature.

Meanwhile, $^{13}$CO and C$^{18}$O emissions lines (Fig. \ref{fig: ALMA 13CO and C18O}) are confined to the central envelope region (5.89- 12.71 km s$^{-1}$ and 7.19-10.45 km s$^{-1}$, respectively), tracing dense, slow-moving material along 
cavity walls. 
Their spatial distribution and lack of high-velocity components suggest these lines trace the surrounding envelope rather than the jet itself, in agreement with earlier envelope studies (e.g., \citealt{2006ApJ...648..504C}).

A multi-wavelength comparison (see Figure \ref{fig: Comparison Image}), overlays ALMA’s $^{12}$CO redshifted outflow, NIRCam F460M emission, and Subaru’s 
H$\alpha$ data. 
The redshifted lobe shows a morphological continuity of the molecular outflow, visible across all wavelengths. 
This indicates that the outflow observed in ALMA and JWST likely continues into the shock front region probed by the Subaru data. 
The NIRCam R32 knot, located at 18.72$\arcsec$ from the central source (see Fig. \ref{fig: JWST Boxes}, Panels A and B), may be associated to this shock emission, which may support the hypothesis that the HH 270 outflow extends well beyond the JWST field, potentially interacting with HH 112, a source previously detected with Spitzer (see Fig. 9 in \citealt{2012AJ....143..106K}) and which lies along the projected path of the HH270 redshifted lobe. This, in turn, could suggest a broader dynamic network of outflows in the L1617 region.

On the blueshifted lobe, NIRCam knots B10-B11 and B12-B13 (see Fig. \ref{fig: JWST Boxes}, Panel D) may be associated to HH 270's optical knots E and D (see Fig. \ref{fig: Comparison Image}), respectively, further reinforcing earlier interpretations of a coherent outflow structure. 
These results support the idea that the SW lobe of HH 270 is being modified downstream, as the leading bowshocks change direction southwards, becoming what is known as HH 110 (\citealt{2012AJ....143..106K}, \citealt{2025ApJ...981...95R}).

The results presented here contribute to our understanding of jet evolution in complex, turbulent star-forming environments. 
The unprecedented spatial resolution of NIRCam has enabled the identification of new knots and cavity structures, while ALMA's molecular line data provides complementary insight into the molecular outflow and its dynamics. 
When combined with optical and infrared observations, the data supports a scenario where the HH 270 outflow is shaped by the environment further downstream from the source. 
The observations do not directly measure the ambient flow proposed by \citealt{2025ApJ...981...95R} driven by HH 451, since they do not cover that region. 
Our findings highlight the value of multi-wavelength, high-resolution observations for disentangling the physical mechanisms driving jet morphology, deflection, and propagation, demonstrating how combining JWST and ALMA data can help unravel jet morphology and molecular outflow structure.

\section{Summary and Conclusion} \label{sec:Conclusion}
We present a multi-wavelength study of the HH 270 system, revealing intricate details of its protostellar outflow and its interaction with the surrounding molecular cloud. 
High-resolution NIRCam images show a detailed bipolar cavity structure and numerous new jet knots in both redshifted and blueshifted lobes. 
ALMA data trace the kinematics of the $^{12}$CO outflow, confirming jet curvature toward the redshifted component and supporting a scenario of jet precession or interaction with dense ambient material. 
The comparison across infrared, optical, and radio wavelengths supports a dynamic picture where HH 270 drives feedback into the molecular cloud and may even interact with nearby HH objects such as HH 112 and HH 110. 
These findings support recent models, including shear-flow-induced jet deflection, offering a refined framework for interpreting jet-ISM interactions. 

The measurements obtained at the spatial scales probed with JWST provide new insights on the physics of the HH 270 protostellar jet. 
In particular, on the asymmetries already imprinted in the outflow collimation process very close to the source and the behavior of molecular gas in its vicinity. 
These observations reveal the detailed morphology of the inner jet, identify new knots within a few hundred au of the central source, and show how the infrared jet emission relates spatially to the surrounding molecular outflow. 
This establishes quantitative benchmarks for future numerical and analytical models of jet-cloud interactions in complex environments. 
Continued high-resolution, multi-epoch observations will be essential to fully resolve the origin of these complex objects and disentangle their temporal evolution, thus understanding the long-term dynamics of systems like HH 270/110.

\begin{acknowledgments}
This research was supported by NASA grant 80NSSC22M0167, and this article uses the following ALMA data: ADS/JAO.ALMA 2018.1.01038.S. 
ALMA is a partnership of ESO (representing its member states), NSF (USA) and NINS (Japan), together with NRC (Canada), MOST and ASIAA (Taiwan), and KASI (Republic of Korea), in cooperation with the Republic of Chile. 
The Joint ALMA Observatory is operated by ESO, AUI/NRAO and NAOJ. The National Radio Astronomy Observatory and Green Bank Observatory are facilities of the U.S. National Science Foundation operated under cooperative agreement by Associated Universities, Inc. 
We extend our gratitude to our collaborators and colleagues for helpful discussion and feedback.
We thank John J. Tobin for providing processed ALMA data products that contributed part of our analysis. 
This research made use of Photutils, an Astropy package for detection and photometry of astronomical sources \citep{larry_bradley_2025_14889440}.
\end{acknowledgments}

\facilities{ALMA:1.3mm, JWST(NIRCam), Subaru Telescope}
\software{\texttt{astroalign} \citep{2020A&C....3200384B}, \texttt{astropy} \citep{2013A&A...558A..33A, 2018AJ....156..123A, 2022ApJ...935..167A}}, CARTA \citep{2021ascl.soft03031C}, CASA \citep{2022PASP..134k4501C}, \texttt{jwst} \citep{2022zndo...7071140B}, \texttt{matplotlib} \citep{2007CSE.....9...90H}, SAOImage DS9 \citep{2000ascl.soft03002S},  \texttt{tclean} \citep{2020PASP..132b4505K}

\bibliography{references}{}

@ARTICLE{2022ApJ...925...39T,
       author = {{Tobin}, John J. and {Offner}, Stella S.~R. and {Kratter}, Kaitlin M. and {Megeath}, S. Thomas and {Sheehan}, Patrick D. and {Looney}, Leslie W. and {Diaz-Rodriguez}, Ana Karla and {Osorio}, Mayra and {Anglada}, Guillem and {Sadavoy}, Sarah I. and {Furlan}, Elise and {Segura-Cox}, Dominique and {Karnath}, Nicole and {van't Hoff}, Merel L.~R. and {van Dishoeck}, Ewine F. and {Li}, Zhi-Yun and {Sharma}, Rajeeb and {Stutz}, Amelia M. and {Tychoniec}, {\L}ukasz},
        title = "{The VLA/ALMA Nascent Disk And Multiplicity (VANDAM) Survey of Orion Protostars. V. A Characterization of Protostellar Multiplicity}",
      journal = {\apj},
         year = 2022,
        month = jan,
       volume = {925},
       number = {1},
          eid = {39},
        pages = {39},
          doi = {10.3847/1538-4357/ac36d2},
archivePrefix = {arXiv},
       eprint = {2111.05801},
 primaryClass = {astro-ph.GA},
       adsurl = {https://ui.adsabs.harvard.edu/abs/2022ApJ...925...39T}
}

@ARTICLE{2020ApJ...890..130T,
       author = {{Tobin}, John J. and {Sheehan}, Patrick D. and {Megeath}, S. Thomas and {D{\'\i}az-Rodr{\'\i}guez}, Ana Karla and {Offner}, Stella S.~R. and {Murillo}, Nadia M. and {van 't Hoff}, Merel L.~R. and {van Dishoeck}, Ewine F. and {Osorio}, Mayra and {Anglada}, Guillem and {Furlan}, Elise and {Stutz}, Amelia M. and {Reynolds}, Nickalas and {Karnath}, Nicole and {Fischer}, William J. and {Persson}, Magnus and {Looney}, Leslie W. and {Li}, Zhi-Yun and {Stephens}, Ian and {Chandler}, Claire J. and {Cox}, Erin and {Dunham}, Michael M. and {Tychoniec}, {\L}ukasz and {Kama}, Mihkel and {Kratter}, Kaitlin and {Kounkel}, Marina and {Mazur}, Brian and {Maud}, Luke and {Patel}, Lisa and {Perez}, Laura and {Sadavoy}, Sarah I. and {Segura-Cox}, Dominique and {Sharma}, Rajeeb and {Stephenson}, Brian and {Watson}, Dan M. and {Wyrowski}, Friedrich},
        title = "{The VLA/ALMA Nascent Disk and Multiplicity (VANDAM) Survey of Orion Protostars. II. A Statistical Characterization of Class 0 and Class I Protostellar Disks}",
      journal = {\apj},
         year = 2020,
        month = feb,
       volume = {890},
       number = {2},
          eid = {130},
        pages = {130},
          doi = {10.3847/1538-4357/ab6f64},
archivePrefix = {arXiv},
       eprint = {2001.04468},
 primaryClass = {astro-ph.GA},
       adsurl = {https://ui.adsabs.harvard.edu/abs/2020ApJ...890..130T}
}

@ARTICLE{2012AJ....143..106K,
       author = {{Kajdi{\v{c}}}, P. and {Reipurth}, B. and {Raga}, A.~C. and {Bally}, J. and {Walawender}, J.},
        title = "{Proper Motions of the HH 110/270 System}",
      journal = {\aj},
         year = 2012,
        month = may,
       volume = {143},
       number = {5},
          eid = {106},
        pages = {106},
          doi = {10.1088/0004-6256/143/5/106},
       adsurl = {https://ui.adsabs.harvard.edu/abs/2012AJ....143..106K}
}

@ARTICLE{2001ARA&A..39..403R,
       author = {{Reipurth}, Bo and {Bally}, John},
        title = "{Herbig-Haro Flows: Probes of Early Stellar Evolution}",
      journal = {\araa},
         year = 2001,
        month = jan,
       volume = {39},
        pages = {403-455},
          doi = {10.1146/annurev.astro.39.1.403},
       adsurl = {https://ui.adsabs.harvard.edu/abs/2001ARA&A..39..403R}
}

@ARTICLE{1998RMxAA..34...69R,
       author = {{Rodr{\'\i}guez}, L.~F. and {Reipurth}, Bo and {Raga}, A.~C. and {Cant{\'o}}, J.},
        title = "{VLA Detection of the Exciting Source of the ``Deflected'' HH 270/110 System}",
      journal = {\rmxaa},
         year = 1998,
        month = oct,
       volume = {34},
        pages = {69-72},
       adsurl = {https://ui.adsabs.harvard.edu/abs/1998RMxAA..34...69R}
}

@ARTICLE{1996A&A...311..989R,
       author = {{Reipurth}, B. and {Raga}, A.~C. and {Heathcote}, S.},
        title = "{HH 110: the grazing collision of a Herbig-Haro flow with a molecular cloud core.}",
      journal = {\aap},
         year = 1996,
        month = jul,
       volume = {311},
        pages = {989-996},
       adsurl = {https://ui.adsabs.harvard.edu/abs/1996A&A...311..989R}
}

@ARTICLE{1996ApJ...462..804N,
       author = {{Noriega-Crespo}, A. and {Garnavich}, P.~M. and {Raga}, A.~C. and {Canto}, J. and {Boehm}, K. -H.},
        title = "{HH 110 Jet Near-Infrared Imaging: The Outflow Mixing Layer}",
      journal = {\apj},
         year = 1996,
        month = may,
       volume = {462},
        pages = {804},
          doi = {10.1086/177195},
       adsurl = {https://ui.adsabs.harvard.edu/abs/1996ApJ...462..804N}
}

@ARTICLE{2020A&C....3200384B,
       author = {{Beroiz}, M. and {Cabral}, J.~B. and {Sanchez}, B.},
        title = "{Astroalign: A Python module for astronomical image registration}",
      journal = {Astronomy and Computing},
         year = 2020,
        month = jul,
       volume = {32},
          eid = {100384},
        pages = {100384},
          doi = {10.1016/j.ascom.2020.100384},
archivePrefix = {arXiv},
       eprint = {1909.02946},
 primaryClass = {astro-ph.IM},
       adsurl = {https://ui.adsabs.harvard.edu/abs/2020A&C....3200384B}
}

@ARTICLE{2005A&A...437..169W,
       author = {{Wang}, H. and {Stecklum}, B. and {Henning}, Th.},
        title = "{New Herbig-Haro objects in the L1617 and L1646 dark clouds}",
      journal = {\aap},
         year = 2005,
        month = jul,
       volume = {437},
       number = {1},
        pages = {169-175},
          doi = {10.1051/0004-6361:20052769},
       adsurl = {https://ui.adsabs.harvard.edu/abs/2005A&A...437..169W}
}

@INCOLLECTION{2008hsf1.book..782R,
       author = {{Reipurth}, B. and {Megeath}, S.~T. and {Bally}, J. and {Walawender}, J.},
        title = "{The L1617 and L1622 Cometary Clouds in Orion}",
    booktitle = {Handbook of Star Forming Regions, Volume I},
         year = 2008,
       editor = {{Reipurth}, B.},
       volume = {4},
        pages = {782},
       adsurl = {https://ui.adsabs.harvard.edu/abs/2008hsf1.book..782R}
}

@ARTICLE{2006ApJ...648..504C,
       author = {{Choi}, Minho and {Tang}, Ya-Wen},
        title = "{Millimeter Imaging of the HH 270 Protostellar Core and Outflow}",
      journal = {\apj},
         year = 2006,
        month = sep,
       volume = {648},
       number = {1},
        pages = {504-509},
          doi = {10.1086/505688},
       adsurl = {https://ui.adsabs.harvard.edu/abs/2006ApJ...648..504C}
}

@ARTICLE{2011A&A...527A..41S,
       author = {{Sep{\'u}lveda}, I. and {Anglada}, G. and {Estalella}, R. and {L{\'o}pez}, R. and {Girart}, J.~M. and {Yang}, J.},
        title = "{Dense gas and the nature of the outflows}",
      journal = {\aap},
         year = 2011,
        month = mar,
       volume = {527},
          eid = {A41},
        pages = {A41},
          doi = {10.1051/0004-6361/200912916},
archivePrefix = {arXiv},
       eprint = {1101.2632},
 primaryClass = {astro-ph.SR},
       adsurl = {https://ui.adsabs.harvard.edu/abs/2011A&A...527A..41S}
}

@ARTICLE{2023Natur.622...48R,
       author = {{Ray}, T.~P. and {McCaughrean}, M.~J. and {Caratti o Garatti}, A. and {Kavanagh}, P.~J. and {Justtanont}, K. and {van Dishoeck}, E.~F. and {Reitsma}, M. and {Beuther}, H. and {Francis}, L. and {Gieser}, C. and {Klaassen}, P. and {Perotti}, G. and {Tychoniec}, L. and {van Gelder}, M. and {Colina}, L. and {Greve}, Th. R. and {G{\"u}del}, M. and {Henning}, Th. and {Lagage}, P.~O. and {{\"O}stlin}, G. and {Vandenbussche}, B. and {Waelkens}, C. and {Wright}, G.},
        title = "{Outflows from the youngest stars are mostly molecular}",
      journal = {\nat},
         year = 2023,
        month = oct,
       volume = {622},
       number = {7981},
        pages = {48-52},
          doi = {10.1038/s41586-023-06551-1},
       adsurl = {https://ui.adsabs.harvard.edu/abs/2023Natur.622...48R}
}

@ARTICLE{2025ApJ...981...95R,
       author = {{Raga}, A.~C. and {Noriega-Crespo}, A. and {Castellanos-Ram{\'\i}rez}, A. and {Cant{\'o}}, J. and {Arce}, H. and {Lebr{\'o}n}, M. and {Morales Ortiz}, J.~L. and {Ortiz Capeles}, A.~N. and {Pantoja}, C.~A.},
        title = "{HH 270/110 as a Jet/Shear Layer Interaction}",
      journal = {\apj},
         year = 2025,
        month = mar,
       volume = {981},
       number = {1},
          eid = {95},
        pages = {95},
          doi = {10.3847/1538-4357/adad6d},
archivePrefix = {arXiv},
       eprint = {2501.13048},
 primaryClass = {astro-ph.SR},
       adsurl = {https://ui.adsabs.harvard.edu/abs/2025ApJ...981...95R}
}

@MISC{2017jwst.prop.1293N,
       author = {{Noriega-Crespo}, Alberto},
        title = "{Collimation zone in Proto-Stellar Jets, the Final Frontier. MIRI MRS and Imager Simultaneous observations}",
 howpublished = {JWST Proposal. Cycle 1, ID. \#1293},
         year = 2017,
        month = aug,
        pages = {1293},
       adsurl = {https://ui.adsabs.harvard.edu/abs/2017jwst.prop.1293N}
}

@software{2000ascl.soft03002S,
       author = {{Smithsonian Astrophysical Observatory}},
        title = "{SAOImage DS9: A utility for displaying astronomical images in the X11 window environment}",
 howpublished = {Astrophysics Source Code Library, record ascl:0003.002},
         year = 2000,
        month = mar,
          eid = {ascl:0003.002},
       adsurl = {https://ui.adsabs.harvard.edu/abs/2000ascl.soft03002S}
}

@software{2021ascl.soft03031C,
       author = {{Comrie}, Angus and {Wang}, Kuo-Song and {Hsu}, Shou-Chieh and {Moraghan}, Anthony and {Harris}, Pamela and {Pang}, Qi and {Pi{\'n}ska}, Adrianna and {Chiang}, Cheng-Chin and {Simmonds}, Rob and {Chang}, Tien-Hao and {Jan}, Hengtai and {Lin}, Ming-Yi},
        title = "{CARTA: Cube Analysis and Rendering Tool for Astronomy}",
 howpublished = {Astrophysics Source Code Library, record ascl:2103.031},
         year = 2021,
        month = mar,
          eid = {ascl:2103.031},
       adsurl = {https://ui.adsabs.harvard.edu/abs/2021ascl.soft03031C}
}

@ARTICLE{2020RMxAA..56...29N,
       author = {{Noriega-Crespo}, A. and {Raga}, A.~C. and {Lora}, V. and {Rodr{\'\i}guez-Ram{\'\i}rez}, J.~C.},
        title = "{The Jet/Counterjet Symmetry of the HH 212 Outflow}",
      journal = {\rmxaa},
         year = 2020,
        month = apr,
       volume = {56},
        pages = {29-38},
          doi = {10.22201/ia.01851101p.2020.56.01.05},
archivePrefix = {arXiv},
       eprint = {1911.12297},
 primaryClass = {astro-ph.IM},
       adsurl = {https://ui.adsabs.harvard.edu/abs/2020RMxAA..56...29N}
}

@ARTICLE{1998AJ....116.2943R,
       author = {{Raga}, A. and {Noriega-Crespo}, A.},
        title = "{A Three-mode, Variable Velocity Jet Model for HH 34}",
      journal = {\aj},
         year = 1998,
        month = dec,
       volume = {116},
       number = {6},
        pages = {2943-2952},
          doi = {10.1086/300641},
archivePrefix = {arXiv},
       eprint = {astro-ph/9808280},
 primaryClass = {astro-ph},
       adsurl = {https://ui.adsabs.harvard.edu/abs/1998AJ....116.2943R}
}

@ARTICLE{2010ApJ...712.1010T,
       author = {{Tobin}, John J. and {Hartmann}, Lee and {Looney}, Leslie W. and {Chiang}, Hsin-Fang},
        title = "{Complex Structure in Class 0 Protostellar Envelopes}",
      journal = {\apj},
         year = 2010,
        month = apr,
       volume = {712},
       number = {2},
        pages = {1010-1028},
          doi = {10.1088/0004-637X/712/2/1010},
archivePrefix = {arXiv},
       eprint = {1002.2362},
 primaryClass = {astro-ph.SR},
       adsurl = {https://ui.adsabs.harvard.edu/abs/2010ApJ...712.1010T}
}

@ARTICLE{2011ApJ...740...45T,
       author = {{Tobin}, John J. and {Hartmann}, Lee and {Chiang}, Hsin-Fang and {Looney}, Leslie W. and {Bergin}, Edwin A. and {Chandler}, Claire J. and {Masqu{\'e}}, Josep M. and {Maret}, S{\'e}bastien and {Heitsch}, Fabian},
        title = "{Complex Structure in Class 0 Protostellar Envelopes. II. Kinematic Structure from Single-dish and Interferometric Molecular Line Mapping}",
      journal = {\apj},
         year = 2011,
        month = oct,
       volume = {740},
       number = {1},
          eid = {45},
        pages = {45},
          doi = {10.1088/0004-637X/740/1/45},
archivePrefix = {arXiv},
       eprint = {1107.4361},
 primaryClass = {astro-ph.SR},
       adsurl = {https://ui.adsabs.harvard.edu/abs/2011ApJ...740...45T}
}

@ARTICLE{2020PASP..132b4505K,
       author = {{Kepley}, Amanda A. and {Tsutsumi}, Takahiro and {Brogan}, Crystal L. and {Indebetouw}, Remy and {Yoon}, Ilsang and {Mason}, Brian and {Donovan Meyer}, Jennifer},
        title = "{Auto-multithresh: A General Purpose Automasking Algorithm}",
      journal = {\pasp},
         year = 2020,
        month = feb,
       volume = {132},
       number = {1008},
          eid = {024505},
        pages = {024505},
          doi = {10.1088/1538-3873/ab5e14},
archivePrefix = {arXiv},
       eprint = {1912.04970},
 primaryClass = {astro-ph.IM},
       adsurl = {https://ui.adsabs.harvard.edu/abs/2020PASP..132b4505K}
}

@ARTICLE{2022PASP..134k4501C,
       author = {{CASA Team} and {Bean}, Ben and {Bhatnagar}, Sanjay and {Castro}, Sandra and {Donovan Meyer}, Jennifer and {Emonts}, Bjorn and {Garcia}, Enrique and {Garwood}, Robert and {Golap}, Kumar and {Gonzalez Villalba}, Justo and {Harris}, Pamela and {Hayashi}, Yohei and {Hoskins}, Josh and {Hsieh}, Mingyu and {Jagannathan}, Preshanth and {Kawasaki}, Wataru and {Keimpema}, Aard and {Kettenis}, Mark and {Lopez}, Jorge and {Marvil}, Joshua and {Masters}, Joseph and {McNichols}, Andrew and {Mehringer}, David and {Miel}, Renaud and {Moellenbrock}, George and {Montesino}, Federico and {Nakazato}, Takeshi and {Ott}, Juergen and {Petry}, Dirk and {Pokorny}, Martin and {Raba}, Ryan and {Rau}, Urvashi and {Schiebel}, Darrell and {Schweighart}, Neal and {Sekhar}, Srikrishna and {Shimada}, Kazuhiko and {Small}, Des and {Steeb}, Jan-Willem and {Sugimoto}, Kanako and {Suoranta}, Ville and {Tsutsumi}, Takahiro and {van Bemmel}, Ilse M. and {Verkouter}, Marjolein and {Wells}, Akeem and {Xiong}, Wei and {Szomoru}, Arpad and {Griffith}, Morgan and {Glendenning}, Brian and {Kern}, Jeff},
        title = "{CASA, the Common Astronomy Software Applications for Radio Astronomy}",
      journal = {\pasp},
         year = 2022,
        month = nov,
       volume = {134},
       number = {1041},
          eid = {114501},
        pages = {114501},
          doi = {10.1088/1538-3873/ac9642},
archivePrefix = {arXiv},
       eprint = {2210.02276},
 primaryClass = {astro-ph.IM},
       adsurl = {https://ui.adsabs.harvard.edu/abs/2022PASP..134k4501C}
}

@ARTICLE{2022ApJ...935..167A,
       author = {{Astropy Collaboration} and {Price-Whelan}, Adrian M. and {Lim}, Pey Lian and {Earl}, Nicholas and {Starkman}, Nathaniel and {Bradley}, Larry and {Shupe}, David L. and {Patil}, Aarya A. and {Corrales}, Lia and {Brasseur}, C.~E. and {N{\"o}the}, Maximilian and {Donath}, Axel and {Tollerud}, Erik and {Morris}, Brett M. and {Ginsburg}, Adam and {Vaher}, Eero and {Weaver}, Benjamin A. and {Tocknell}, James and {Jamieson}, William and {van Kerkwijk}, Marten H. and {Robitaille}, Thomas P. and {Merry}, Bruce and {Bachetti}, Matteo and {G{\"u}nther}, H. Moritz and {Aldcroft}, Thomas L. and {Alvarado-Montes}, Jaime A. and {Archibald}, Anne M. and {B{\'o}di}, Attila and {Bapat}, Shreyas and {Barentsen}, Geert and {Baz{\'a}n}, Juanjo and {Biswas}, Manish and {Boquien}, M{\'e}d{\'e}ric and {Burke}, D.~J. and {Cara}, Daria and {Cara}, Mihai and {Conroy}, Kyle E. and {Conseil}, Simon and {Craig}, Matthew W. and {Cross}, Robert M. and {Cruz}, Kelle L. and {D'Eugenio}, Francesco and {Dencheva}, Nadia and {Devillepoix}, Hadrien A.~R. and {Dietrich}, J{\"o}rg P. and {Eigenbrot}, Arthur Davis and {Erben}, Thomas and {Ferreira}, Leonardo and {Foreman-Mackey}, Daniel and {Fox}, Ryan and {Freij}, Nabil and {Garg}, Suyog and {Geda}, Robel and {Glattly}, Lauren and {Gondhalekar}, Yash and {Gordon}, Karl D. and {Grant}, David and {Greenfield}, Perry and {Groener}, Austen M. and {Guest}, Steve and {Gurovich}, Sebastian and {Handberg}, Rasmus and {Hart}, Akeem and {Hatfield-Dodds}, Zac and {Homeier}, Derek and {Hosseinzadeh}, Griffin and {Jenness}, Tim and {Jones}, Craig K. and {Joseph}, Prajwel and {Kalmbach}, J. Bryce and {Karamehmetoglu}, Emir and {Ka{\l}uszy{\'n}ski}, Miko{\l}aj and {Kelley}, Michael S.~P. and {Kern}, Nicholas and {Kerzendorf}, Wolfgang E. and {Koch}, Eric W. and {Kulumani}, Shankar and {Lee}, Antony and {Ly}, Chun and {Ma}, Zhiyuan and {MacBride}, Conor and {Maljaars}, Jakob M. and {Muna}, Demitri and {Murphy}, N.~A. and {Norman}, Henrik and {O'Steen}, Richard and {Oman}, Kyle A. and {Pacifici}, Camilla and {Pascual}, Sergio and {Pascual-Granado}, J. and {Patil}, Rohit R. and {Perren}, Gabriel I. and {Pickering}, Timothy E. and {Rastogi}, Tanuj and {Roulston}, Benjamin R. and {Ryan}, Daniel F. and {Rykoff}, Eli S. and {Sabater}, Jose and {Sakurikar}, Parikshit and {Salgado}, Jes{\'u}s and {Sanghi}, Aniket and {Saunders}, Nicholas and {Savchenko}, Volodymyr and {Schwardt}, Ludwig and {Seifert-Eckert}, Michael and {Shih}, Albert Y. and {Jain}, Anany Shrey and {Shukla}, Gyanendra and {Sick}, Jonathan and {Simpson}, Chris and {Singanamalla}, Sudheesh and {Singer}, Leo P. and {Singhal}, Jaladh and {Sinha}, Manodeep and {Sip{\H{o}}cz}, Brigitta M. and {Spitler}, Lee R. and {Stansby}, David and {Streicher}, Ole and {{\v{S}}umak}, Jani and {Swinbank}, John D. and {Taranu}, Dan S. and {Tewary}, Nikita and {Tremblay}, Grant R. and {de Val-Borro}, Miguel and {Van Kooten}, Samuel J. and {Vasovi{\'c}}, Zlatan and {Verma}, Shresth and {de Miranda Cardoso}, Jos{\'e} Vin{\'\i}cius and {Williams}, Peter K.~G. and {Wilson}, Tom J. and {Winkel}, Benjamin and {Wood-Vasey}, W.~M. and {Xue}, Rui and {Yoachim}, Peter and {Zhang}, Chen and {Zonca}, Andrea and {Astropy Project Contributors}},
        title = "{The Astropy Project: Sustaining and Growing a Community-oriented Open-source Project and the Latest Major Release (v5.0) of the Core Package}",
      journal = {\apj},
         year = 2022,
        month = aug,
       volume = {935},
       number = {2},
          eid = {167},
        pages = {167},
          doi = {10.3847/1538-4357/ac7c74},
archivePrefix = {arXiv},
       eprint = {2206.14220},
 primaryClass = {astro-ph.IM},
       adsurl = {https://ui.adsabs.harvard.edu/abs/2022ApJ...935..167A}
}

@software{2022zndo...7071140B,
       author = {{Bushouse}, Howard and {Eisenhamer}, Jonathan and {Dencheva}, Nadia and {Davies}, James and {Greenfield}, Perry and {Morrison}, Jane and {Hodge}, Phil and {Simon}, Bernie and {Grumm}, David and {Droettboom}, Michael and {Slavich}, Edward and {Sosey}, Megan and {Pauly}, Tyler and {Miller}, Todd and {Jedrzejewski}, Robert and {Hack}, Warren and {Davis}, David and {Crawford}, Steven and {Law}, David and {Gordon}, Karl and {Regan}, Michael and {Cara}, Mihai and {MacDonald}, Ken and {Bradley}, Larry and {Shanahan}, Clare and {Jamieson}, William and {Teodoro}, Mairan and {Williams}, Thomas},
        title = "{JWST Calibration Pipeline}",
         year = 2022,
        month = sep,
          eid = {10.5281/zenodo.7071140},
          doi = {10.5281/zenodo.7071140},
      version = {1.7.0},
    publisher = {Zenodo},
       adsurl = {https://ui.adsabs.harvard.edu/abs/2022zndo...7071140B}
}

@ARTICLE{2007CSE.....9...90H,
       author = {{Hunter}, John D.},
        title = "{Matplotlib: A 2D Graphics Environment}",
      journal = {Computing in Science and Engineering},
         year = 2007,
        month = jan,
       volume = {9},
       number = {3},
        pages = {90-95},
          doi = {10.1109/MCSE.2007.55},
       adsurl = {https://ui.adsabs.harvard.edu/abs/2007CSE.....9...90H}
}

@ARTICLE{2018AJ....156..123A,
       author = {{Astropy Collaboration} and {Price-Whelan}, A.~M. and {Sip{\H{o}}cz}, B.~M. and {G{\"u}nther}, H.~M. and {Lim}, P.~L. and {Crawford}, S.~M. and {Conseil}, S. and {Shupe}, D.~L. and {Craig}, M.~W. and {Dencheva}, N. and {Ginsburg}, A. and {VanderPlas}, J.~T. and {Bradley}, L.~D. and {P{\'e}rez-Su{\'a}rez}, D. and {de Val-Borro}, M. and {Aldcroft}, T.~L. and {Cruz}, K.~L. and {Robitaille}, T.~P. and {Tollerud}, E.~J. and {Ardelean}, C. and {Babej}, T. and {Bach}, Y.~P. and {Bachetti}, M. and {Bakanov}, A.~V. and {Bamford}, S.~P. and {Barentsen}, G. and {Barmby}, P. and {Baumbach}, A. and {Berry}, K.~L. and {Biscani}, F. and {Boquien}, M. and {Bostroem}, K.~A. and {Bouma}, L.~G. and {Brammer}, G.~B. and {Bray}, E.~M. and {Breytenbach}, H. and {Buddelmeijer}, H. and {Burke}, D.~J. and {Calderone}, G. and {Cano Rodr{\'\i}guez}, J.~L. and {Cara}, M. and {Cardoso}, J.~V.~M. and {Cheedella}, S. and {Copin}, Y. and {Corrales}, L. and {Crichton}, D. and {D'Avella}, D. and {Deil}, C. and {Depagne}, {\'E}. and {Dietrich}, J.~P. and {Donath}, A. and {Droettboom}, M. and {Earl}, N. and {Erben}, T. and {Fabbro}, S. and {Ferreira}, L.~A. and {Finethy}, T. and {Fox}, R.~T. and {Garrison}, L.~H. and {Gibbons}, S.~L.~J. and {Goldstein}, D.~A. and {Gommers}, R. and {Greco}, J.~P. and {Greenfield}, P. and {Groener}, A.~M. and {Grollier}, F. and {Hagen}, A. and {Hirst}, P. and {Homeier}, D. and {Horton}, A.~J. and {Hosseinzadeh}, G. and {Hu}, L. and {Hunkeler}, J.~S. and {Ivezi{\'c}}, {\v{Z}}. and {Jain}, A. and {Jenness}, T. and {Kanarek}, G. and {Kendrew}, S. and {Kern}, N.~S. and {Kerzendorf}, W.~E. and {Khvalko}, A. and {King}, J. and {Kirkby}, D. and {Kulkarni}, A.~M. and {Kumar}, A. and {Lee}, A. and {Lenz}, D. and {Littlefair}, S.~P. and {Ma}, Z. and {Macleod}, D.~M. and {Mastropietro}, M. and {McCully}, C. and {Montagnac}, S. and {Morris}, B.~M. and {Mueller}, M. and {Mumford}, S.~J. and {Muna}, D. and {Murphy}, N.~A. and {Nelson}, S. and {Nguyen}, G.~H. and {Ninan}, J.~P. and {N{\"o}the}, M. and {Ogaz}, S. and {Oh}, S. and {Parejko}, J.~K. and {Parley}, N. and {Pascual}, S. and {Patil}, R. and {Patil}, A.~A. and {Plunkett}, A.~L. and {Prochaska}, J.~X. and {Rastogi}, T. and {Reddy Janga}, V. and {Sabater}, J. and {Sakurikar}, P. and {Seifert}, M. and {Sherbert}, L.~E. and {Sherwood-Taylor}, H. and {Shih}, A.~Y. and {Sick}, J. and {Silbiger}, M.~T. and {Singanamalla}, S. and {Singer}, L.~P. and {Sladen}, P.~H. and {Sooley}, K.~A. and {Sornarajah}, S. and {Streicher}, O. and {Teuben}, P. and {Thomas}, S.~W. and {Tremblay}, G.~R. and {Turner}, J.~E.~H. and {Terr{\'o}n}, V. and {van Kerkwijk}, M.~H. and {de la Vega}, A. and {Watkins}, L.~L. and {Weaver}, B.~A. and {Whitmore}, J.~B. and {Woillez}, J. and {Zabalza}, V. and {Astropy Contributors}},
        title = "{The Astropy Project: Building an Open-science Project and Status of the v2.0 Core Package}",
      journal = {\aj},
         year = 2018,
        month = sep,
       volume = {156},
       number = {3},
          eid = {123},
        pages = {123},
          doi = {10.3847/1538-3881/aabc4f},
archivePrefix = {arXiv},
       eprint = {1801.02634},
 primaryClass = {astro-ph.IM},
       adsurl = {https://ui.adsabs.harvard.edu/abs/2018AJ....156..123A}
}

@ARTICLE{2013A&A...558A..33A,
       author = {{Astropy Collaboration} and {Robitaille}, Thomas P. and {Tollerud}, Erik J. and {Greenfield}, Perry and {Droettboom}, Michael and {Bray}, Erik and {Aldcroft}, Tom and {Davis}, Matt and {Ginsburg}, Adam and {Price-Whelan}, Adrian M. and {Kerzendorf}, Wolfgang E. and {Conley}, Alexander and {Crighton}, Neil and {Barbary}, Kyle and {Muna}, Demitri and {Ferguson}, Henry and {Grollier}, Fr{\'e}d{\'e}ric and {Parikh}, Madhura M. and {Nair}, Prasanth H. and {Unther}, Hans M. and {Deil}, Christoph and {Woillez}, Julien and {Conseil}, Simon and {Kramer}, Roban and {Turner}, James E.~H. and {Singer}, Leo and {Fox}, Ryan and {Weaver}, Benjamin A. and {Zabalza}, Victor and {Edwards}, Zachary I. and {Azalee Bostroem}, K. and {Burke}, D.~J. and {Casey}, Andrew R. and {Crawford}, Steven M. and {Dencheva}, Nadia and {Ely}, Justin and {Jenness}, Tim and {Labrie}, Kathleen and {Lim}, Pey Lian and {Pierfederici}, Francesco and {Pontzen}, Andrew and {Ptak}, Andy and {Refsdal}, Brian and {Servillat}, Mathieu and {Streicher}, Ole},
        title = "{Astropy: A community Python package for astronomy}",
      journal = {\aap},
         year = 2013,
        month = oct,
       volume = {558},
          eid = {A33},
        pages = {A33},
          doi = {10.1051/0004-6361/201322068},
archivePrefix = {arXiv},
       eprint = {1307.6212},
 primaryClass = {astro-ph.IM},
       adsurl = {https://ui.adsabs.harvard.edu/abs/2013A&A...558A..33A}
}

@software{larry_bradley_2025_14889440,
  author       = {Larry Bradley and
                  Brigitta Sip{\H o}cz and
                  Thomas Robitaille and
                  Erik Tollerud and
                  Z\`e Vin{\'{\i}}cius and
                  Christoph Deil and
                  Kyle Barbary and
                  Tom J Wilson and
                  Ivo Busko and
                  Axel Donath and
                  Hans Moritz G{\"u}nther and
                  Mihai Cara and
                  P. L. Lim and
                  Sebastian Me{\ss}linger and
                  Zach Burnett and
                  Simon Conseil and
                  Michael Droettboom and
                  Azalee Bostroem and
                  E. M. Bray and
                  Lars Andersen Bratholm and
                  William Jamieson and
                  Adam Ginsburg and
                  Geert Barentsen and
                  Matt Craig and
                  Sergio Pascual and
                  Shivangee Rathi and
                  Marshall Perrin and
                  Brett M. Morris},
  title        = {astropy/photutils: 2.2.0},
  month        = feb,
  year         = 2025,
  publisher    = {Zenodo},
  version      = {2.2.0},
  doi          = {10.5281/zenodo.14889440},
  url          = {https://doi.org/10.5281/zenodo.14889440},
  swhid        = {swh:1:dir:11159107f27a28985192ed1118b1f2055709d093
                   ;origin=https://doi.org/10.5281/zenodo.596036;visi
                   t=swh:1:snp:ae8c4a55d349d43e53cfe9ce92e678fcfe840f
                   3b;anchor=swh:1:rel:0117f67e8888adcdfc85308287dd9c
                   854b466389;path=astropy-photutils-ffb96c5
                  },
}

@ARTICLE{2010MNRAS.406.2193L,
       author = {{L{\'o}pez}, R. and {Garc{\'\i}a-Lorenzo}, B. and {S{\'a}nchez}, S.~F. and {G{\'o}mez}, G. and {Estalella}, R. and {Riera}, A.},
        title = "{The complex structure of HH 110 as revealed from Integral Field Spectroscopy}",
      journal = {\mnras},
         year = 2010,
        month = aug,
       volume = {406},
       number = {4},
        pages = {2193-2205},
          doi = {10.1111/j.1365-2966.2010.16831.x},
archivePrefix = {arXiv},
       eprint = {1004.2123},
 primaryClass = {astro-ph.GA},
       adsurl = {https://ui.adsabs.harvard.edu/abs/2010MNRAS.406.2193L}
}

@ARTICLE{2024ApJ...966..225A,
       author = {{Alarc{\'o}n}, Felipe and {Bergin}, Edwin A. and {Cugno}, Gabriele},
        title = "{Extinction Values toward Embedded Planets in Protoplanetary Disks Estimated from Hydrodynamic Simulations}",
      journal = {\apj},
         year = 2024,
        month = may,
       volume = {966},
       number = {2},
          eid = {225},
        pages = {225},
          doi = {10.3847/1538-4357/ad3938},
archivePrefix = {arXiv},
       eprint = {2404.05788},
 primaryClass = {astro-ph.EP},
       adsurl = {https://ui.adsabs.harvard.edu/abs/2024ApJ...966..225A}
}

@ARTICLE{1997AJ....114..757R,
       author = {{Reipurth}, Bo and {Hartigan}, Patrick and {Heathcote}, Steve and {Morse}, Jon A. and {Bally}, John},
        title = "{Hubble Space Telescope Images of the HH 111 Jet.}",
      journal = {\aj},
         year = 1997,
        month = aug,
       volume = {114},
        pages = {757-780},
          doi = {10.1086/118509},
       adsurl = {https://ui.adsabs.harvard.edu/abs/1997AJ....114..757R}
}

@ARTICLE{2002ApJ...568..733M,
       author = {{Masciadri}, E. and {Raga}, A.~C.},
        title = "{Herbig-Haro Jets from Orbiting Sources}",
      journal = {\apj},
         year = 2002,
        month = apr,
       volume = {568},
       number = {2},
        pages = {733-742},
          doi = {10.1086/338767},
       adsurl = {https://ui.adsabs.harvard.edu/abs/2002ApJ...568..733M}
}

@ARTICLE{2012AJ....144...61E,
       author = {{Estalella}, Robert and {L{\'o}pez}, Rosario and {Anglada}, Guillem and {G{\'o}mez}, Gabriel and {Riera}, Angels and {Carrasco-Gonz{\'a}lez}, Carlos},
        title = "{The Counterjet of HH 30: New Light on Its Binary Driving Source}",
      journal = {\aj},
         year = 2012,
        month = aug,
       volume = {144},
       number = {2},
          eid = {61},
        pages = {61},
          doi = {10.1088/0004-6256/144/2/61},
archivePrefix = {arXiv},
       eprint = {1206.3391},
 primaryClass = {astro-ph.GA},
       adsurl = {https://ui.adsabs.harvard.edu/abs/2012AJ....144...61E}
}

@ARTICLE{2007AJ....133.2799A,
       author = {{Anglada}, Guillem and {L{\'o}pez}, Rosario and {Estalella}, Robert and {Masegosa}, Josefa and {Riera}, Angels and {Raga}, Alejandro C.},
        title = "{Proper Motions of the Jets in the Region of HH 30 and HL/XZ Tau: Evidence for a Binary Exciting Source of the HH 30 Jet}",
      journal = {\aj},
         year = 2007,
        month = jun,
       volume = {133},
       number = {6},
        pages = {2799-2814},
          doi = {10.1086/517493},
archivePrefix = {arXiv},
       eprint = {astro-ph/0703155},
 primaryClass = {astro-ph},
       adsurl = {https://ui.adsabs.harvard.edu/abs/2007AJ....133.2799A}
}

@ARTICLE{2023PASP..135b8001R,
       author = {{Rieke}, Marcia J. and {Kelly}, Douglas M. and {Misselt}, Karl and {Stansberry}, John and {Boyer}, Martha and {Beatty}, Thomas and {Egami}, Eiichi and {Florian}, Michael and {Greene}, Thomas P. and {Hainline}, Kevin and {Leisenring}, Jarron and {Roellig}, Thomas and {Schlawin}, Everett and {Sun}, Fengwu and {Tinnin}, Lee and {Williams}, Christina C. and {Willmer}, Christopher N.~A. and {Wilson}, Debra and {Clark}, Charles R. and {Rohrbach}, Scott and {Brooks}, Brian and {Canipe}, Alicia and {Correnti}, Matteo and {DiFelice}, Audrey and {Gennaro}, Mario and {Girard}, Julien H. and {Hartig}, George and {Hilbert}, Bryan and {Koekemoer}, Anton M. and {Nikolov}, Nikolay K. and {Pirzkal}, Norbert and {Rest}, Armin and {Robberto}, Massimo and {Sunnquist}, Ben and {Telfer}, Randal and {Wu}, Chi Rai and {Ferry}, Malcolm and {Lewis}, Dan and {Baum}, Stefi and {Beichman}, Charles and {Doyon}, Ren{\'e} and {Dressler}, Alan and {Eisenstein}, Daniel J. and {Ferrarese}, Laura and {Hodapp}, Klaus and {Horner}, Scott and {Jaffe}, Daniel T. and {Johnstone}, Doug and {Krist}, John and {Martin}, Peter and {McCarthy}, Donald W. and {Meyer}, Michael and {Rieke}, George H. and {Trauger}, John and {Young}, Erick T.},
        title = "{Performance of NIRCam on JWST in Flight}",
      journal = {\pasp},
         year = 2023,
        month = feb,
       volume = {135},
       number = {1044},
          eid = {028001},
        pages = {028001},
          doi = {10.1088/1538-3873/acac53},
archivePrefix = {arXiv},
       eprint = {2212.12069},
 primaryClass = {astro-ph.IM},
       adsurl = {https://ui.adsabs.harvard.edu/abs/2023PASP..135b8001R}
}

@ARTICLE{2015GeoJI.200..111W,
       author = {{Wang}, Yanghua},
        title = "{The Ricker wavelet and the Lambert W function}",
      journal = {Geophysical Journal International},
         year = 2015,
        month = jan,
       volume = {200},
       number = {1},
        pages = {111-115},
          doi = {10.1093/gji/ggu384},
       adsurl = {https://ui.adsabs.harvard.edu/abs/2015GeoJI.200..111W}
}
\bibliographystyle{aasjournalv7}
\end{document}